\documentclass[10pt,journal, compsoc]{IEEEtran}
\usepackage{amsmath,amsfonts}
\usepackage{algorithmic}
\usepackage{algorithm}
\usepackage{array}
\usepackage[caption=false,font=normalsize,labelfont=sf,textfont=sf]{subfig}
\usepackage{textcomp}
\usepackage{stfloats}
\usepackage{url}
\usepackage{verbatim}
\usepackage{graphicx}
\usepackage{cite}
\usepackage{adjustbox}
\usepackage{enumitem}
\usepackage{multicol, multirow}
\usepackage{soul}
\usepackage[table]{xcolor}
\usepackage{colortbl}
\usepackage[table]{xcolor}
\usepackage{longtable}
\usepackage{adjustbox}
\usepackage{tcolorbox}
\usepackage{lipsum}

\usepackage{hyperref}

\newcolumntype{L}{>{\raggedright\arraybackslash}p{0.15\linewidth}}
\newcolumntype{M}{>{\raggedright\arraybackslash}p{0.8\linewidth}}
\newcolumntype{D}{>{\raggedright\arraybackslash}p{0.1\linewidth}}
\newcolumntype{R}{>{\raggedright\arraybackslash}p{0.35\linewidth}}

\definecolor{rowcolor}{gray}{0.95} 

\begin{document}

\title{\textbf{SoK: Leveraging Transformers for Malware Analysis}}


\author{Pradip Kunwar\(^1\), 
Kshitiz Aryal\(^1\), Maanak Gupta\(^1\),~\IEEEmembership{Senior Member, IEEE},
Mahmoud Abdelsalam\(^2\), Elisa Bertino\(^3\),~\IEEEmembership{Fellow, IEEE} \\
\textit{\(^1\)Tennessee Tech University, \(^2\)North Carolina A\&T State University, \(^3\)Purdue University}}



\maketitle

\begin{abstract}
The introduction of transformers has been an important breakthrough for AI research and application, as transformers are the foundation behind Generative AI. 
Transformers are promising in cybersecurity, especially malware analysis. The reason is the flexibility of the transformer models in handling long sequential features and understanding contextual relationships. 
However, as the use of transformers for malware analysis 
is still in the infancy stage, it is critical to evaluate, systematize, and contextualize existing literature to foster future research.
This Systematization of Knowledge (SoK) paper aims to provide a comprehensive analysis of transformer-based approaches designed for malware analysis. Based on our systematic analysis of existing knowledge, we structure and propose taxonomies based on: (a) how different transformers are adapted, organized, and modified across various use cases; and (b) how diverse feature types and their representation capabilities are reflected. We also provide an inventory of datasets used to explore multiple research avenues in the use of transformers for malware analysis and discuss open challenges with future research directions. We believe that this SoK paper will assist the research community in gaining detailed insights from existing work and will serve as a foundational resource for implementing novel research using transformers for malware analysis. 


\end{abstract}


\input{}

\section{Introduction}
\IEEEPARstart{T}he growing dependency on digital technology and connectivity has led to an unprecedented surge in the generation and distribution of malicious software, also referred to as malware~\cite{avtestmalwareStats}. Malware poses significant threats to cybersecurity, targeting individuals, organizations, and critical infrastructures in various forms, including viruses, worms, ransomware, etc. As malware evolves in complexity~\cite{jang2014survey,aryal2021survey}, improving classification, detection, and mitigation methods is critical.
To address this growing challenge in malware analysis, Artificial Intelligence (AI) and Machine learning (ML) based solutions, including Deep Learning (DL)~\cite{gopinath2023comprehensive}, Computer Vision~\cite{raff2018malware, abijah2020vision}, and Natural Language Processing (NLP)~\cite{mimura2022applying, tran2017nlp}, have been recently proposed.
These techniques help automate feature extraction and improve static and dynamic analysis for detection/classification tasks against sophisticated and evasive malware. Among various approaches, transformer-based models are recently garnering significant attention. The transformer model introduced by Vaswani et al.~\cite{vaswani2017attention} has revolutionized a broad range of tasks~\cite{lin2022survey} in various domains including 
NLP~\cite{bracsoveanu2020visualizing, gillioz2020overview} for text classification~\cite{park2022efficient}, machine translation, 
question answering~\cite{kenton2019bert} and text generation~\cite{achiam2023gpt}; computer vision for image classification~\cite{dosovitskiy2020image, khan2022transformers, han2022survey}, object detection \cite{carion2020end}, image and video generation~\cite{ramesh2022hierarchical, videoworldsimulators2024}; speech recognition~\cite{dong2018speech}, and etc. with significant performance improvements. 


In malware analysis, transformers have begun to emerge as a versatile and powerful tool~\cite{li2021mad,fan2021heterogeneous}. Transformers excel in capturing intricate patterns, such as spatial, temporal, structural, etc., across high-dimensional data making them well-suited for
complex static and dynamic malware analysis. These models analyze extensive datasets from raw binary files, disassembled codes, graphs, images, and many other feature representations of malware to uncover malicious behavior and identify emerging threats. Furthermore, their application spans various malware analysis sub-domains including detection~\cite{dehunting,fan2021heterogeneous,lu2022research,li2023iot,ravi2023vit4mal} and classification\cite{or2021pay,rahali2023malbertv2}, binary code similarity detection~\cite{li2023gental}, evasion techniques~\cite{hu2021single}, as well as explanation and 
interpretation~\cite{li2021mad, ullah2022explainable}, underlining their capability to understand malicious behavior accurately.

Although transformers were first introduced in 2017, their integration into malware analysis is relatively at an initial stage, presenting opportunities for future research and advancements. Therefore, it is critical to organize the existing knowledge based on its application, novel research findings, empirical studies, and analysis, along with open challenges to understand and foster future research leveraging transformers in malware analysis. In this paper, we explore the evolution of transformers, their impact on malware analysis, and highlight various approaches and challenges encountered by the research community. Through an extensive literature review, we examine and systematize the knowledge based on the use cases of transformers, their types, modeling aspects, representation techniques, and different features employed in adapting transformer architectures to address challenges in malware analysis. Further, to solve the issue of the lack of a comprehensive dataset repository, we compile an inventory of features, representation techniques, and datasets to assist the community.


\textbf{Methodology and Literature Search}:
We collected available literature from several sources that cover a broad range of research using transformer or its modified version for malware analysis tasks. Along with relevant conference venues and journals, we carried out searches on platforms such as  Google Scholar and Research Rabbit using keywords like \textit{transformers in malware analysis}, \textit{malware transformer}, \textit{transformer survey}, \textit{transformer malware survey}, \textit{sok transformer malware}, \textit{attention malware}, etc. We collected 43 papers covering the period from 2017 to 2023, focusing on the application of transformers for malware analysis. 
From the collected articles, we extracted 30 different 
categories of information, covering objectives, motivations, end goals, domains, features, datasets, representation techniques, modeling, transformer type, need for transformer, weakness, challenges, gaps, future scopes, etc. We assessed the extracted information multiple times to critically analyze and structure the knowledge.

We synthesized and structured the available knowledge based on five major aspects by finding the answers to the questions listed below:
\begin{itemize}[leftmargin=*]
    \item \textbf{Purposes of use for transformers}: Why are the transformers used in malware analysis, and what are the similarities in those use cases?
    \item \textbf{End goals}: Besides the purposes for using transformers, what are the different objectives, such as malware detection, malware classification, detection evasion, etc.?
    \item \textbf{Type of transformer}: Which transformers architectures are used to analyze malware? Are there custom enhancements or modifications performed in the existing variations of transformer architectures?
    \item \textbf{Features}: What are the different features, representing malware behaviors, used to accomplish the end goals that align with the working mechanism of transformers?
    \item \textbf{Datasets}: What are the different datasets used in the malware analysis community that provide the features to model the representation using the transformers?
\end{itemize}

The remainder of this paper is organized as follows. Section~\ref{Background} introduces basic concepts about malware analysis, and the transformer architecture and its evolution. 
Section~\ref{Systematization: transformers in malware Analysis} presents the systematization of transformers for malware analysis followed by Section~\ref{sec:future} discussing the limitations, open challenges, and future directions. Section~\ref{sec:summary} summarizes and concludes the paper. Appendix Section 6 presents in-depth technical discussion on specific transformer architectures and their modifications, a comparative analysis of the performance of different transformer models across various malware tasks, and practical insights on the implementation of transformer models in real-world systems.

\vspace{-1.5mm}

\section{Background}
\label{Background}
\subsection{\textbf{Malware Analysis}}
\label{malware Analysis}
In general, malware analysis techniques examine malicious software and unwanted code in computer systems or networks. This analysis helps to understand behaviors, intents, functions, and impacts of malicious code to assist in anticipating future activities and mitigating attacks. Traditionally, malware analysis is divided into three approaches: static, dynamic, and hybrid. Static analysis is performed without actually executing a program, dynamic analysis requires the program's execution in an isolated virtual environment (i.e., a sandbox), and the hybrid approach combines both approaches. 

Static analysis is based on signature-based approach~\cite{zheng2013droid} and offline code examination such as memory corruption flaws~\cite{chen2004model}. However, maintaining the signatures database is impractical because of exponentially increasing malware volumes and the inability to recognize zero-day malware. In addition, static techniques fail to detect evolving malware that integrates evasive techniques such as obfuscation and polymorphism. To mitigate such limitations, dynamic analysis is used, which examines malware interactions with the operating system, memory, network, and applications, thus tracking its actions, network communications, and modifications to system files and the registry~\cite{damodaran2017comparison}. To leverage both benefits, the hybrid approach combines static and dynamic analysis to gather critical features for performing different malware analysis tasks. However, applying conventional methods is less effective in detecting evolving and sophisticated zero-day malware~\cite{djenna2023artificial}. In addition, advanced malware typically can detect the presence of a sandbox environment and, in turn, can evade behavior-based detection approaches by ceasing or altering their true malicious behavior. A significant limitation of conventional approaches
is their inability to generalize from the known patterns to detect new, sophisticated threats that have not been previously encountered, which results in a higher rate of false negatives~\cite{caviglione2020tight}.

To address these issues, the application of 
ML in malware analysis has increased significantly to enhance real-time detection and classification with high accuracy and low false positives~\cite{li2021mad, dehunting,kimmel2021recurrent,kimmell2021analyzing}. ML and in particular deep learning (DL) models can automate the detection of new and evolving malware variants by learning and distinguishing complex patterns from large datasets of benign and malicious samples~\cite{gibert2020rise}. These practices enhance the detection of patterns, anomalies, behaviors, and intents indicative of malware, improving the speed and accuracy of detection and response.
With the ML advancements, adopting such models has offered significant possibilities for new approaches and techniques to strengthen malware analysis techniques. Among these advancements, the transformer architecture stands out for its ability to handle sequential data, making it particularly suitable for analyzing malware's complex and polymorphic nature.
\vspace{-1.5mm}

\subsection{\textbf{Transformer}}
\label{transformer}
In this section we discuss the transformer architecture, tracing its development and historical knowledge of its evolution over time, and present how the development progressed to solve contemporary issues. 
\subsubsection{\textbf{Evolution of Transformer Architecture}}
\begin{figure*}
    \centering
    \captionsetup{justification=centering,margin=0cm}
    \includegraphics[width=\textwidth]
    {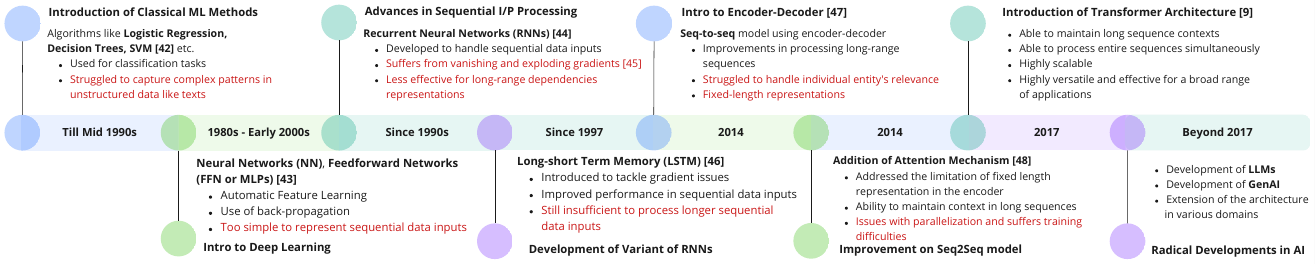}
    \caption{Evolution of the Transformer Architecture: A Milestone in Advancing Machine Translation Task \textit{(Note: Texts in red highlight the shortcomings)}}
    \label{fig:timeline}
     \vspace{-3mm}
\end{figure*}

The development of the transformer model showcases the decades of progress in ML. 
As shown in Fig.~\ref{fig:timeline},  it can all be traced back to the introduction of classical ML algorithms like Logistic Regression, Decision Trees, 
and SVMs~\cite{cortes1995support}, which laid the groundwork for understanding structured data. However, these methods struggled with representing sequential and temporal data, which were critical for machine translation tasks—the very solutions the research community was eagerly seeking. As neural networks, feed-forward networks, and Multi-Layer Perceptron (MLP) equipped with backpropagation~\cite{rumelhart1986learning} were introduced, these DL 
methods could learn complex and nonlinear relationships. But, the models were too simple to represent contextual semantics. Thus, the Recurrent Neural Networks (RNNs)~\cite{elman1990finding} were developed to introduce memory in the network but faced significant challenges in learning long-term dependencies due to vanishing and exploding gradient issues~\cite{hochreiter1998vanishing}. The introduction of Long Short-Term Memory (LSTM) partially addressed 
this issue~\cite{hochreiter1997long} with better memory mechanisms. Also, the encoder-decoder architecture~\cite{sutskever2014sequence} and subsequent integration of the attention mechanism in the 
seq2seq model~\cite{bahdanau2014neural} further enhanced the network’s capabilities to improve attention to the relevant parts of the input sequence. Later, in 2017, Vaswani et al.~\cite{vaswani2017attention} proposed the revolutionary transformer architecture, which offered abilities like scalability, versatility, unprecedented parallelization, and efficiency with performance improvements. 

Fig.~\ref{fig:timeline} illustrates the evolution of transformer architecture through a series of enhancements to existing machine translation solutions. However, this groundbreaking development not only revolutionized machine translation tasks but also extended its application to a wide range of sequential tasks across various fields, including NLP~\cite{bracsoveanu2020visualizing,gillioz2020overview}, computer vision~\cite{khan2022transformers,han2022survey,carion2020end}, and cybersecurity~\cite{li2021mad,fan2021heterogeneous}. While the timeline depicted in the figure extends only until 2017, 
it is important to note that there have been significant developments since then. In the fields of Generative AI and Large Language Models, the expanded applications of transformers, groundbreaking innovations, such as GPT-4 Chatbot~\cite{achiam2023gpt}, Sora video generator model~\cite{videoworldsimulators2024}, DALL-E image generator~\cite{ramesh2022hierarchical} and beyond, highlight their continued impact and versatility in shaping AI. 

\vspace{-2.5mm}
\subsubsection{\textbf{Transformer Architecture}}
A standard transformer architecture, also known as `Vanilla transformer', consists of two sub-networks called encoder and  decoder, as shown in Fig.~\ref{fig:transformer}. In general, the encoder network maps the input sequence into contextual representations which are further processed by the decoder network to generate an output sequence one at a time in an auto-regressive manner i.e. the current value in a sequence is processed as a function of its all previous values.

\textit{\textbf{Input Sequence:}}
For a machine-translation task, the encoder's input sequence is the language to be translated and the first input token for the decoder is the null character, which starts the translation of the language. Once the first token is generated, the output token is shifted right and amended to the existing input sequence and sent to the decoder in an autoregressive manner. These inputs are then embedded 
(as shown in Fig.~\ref{fig:transformer}) with dimension \textit{d} before being passed to the positional encoding layer. With these different inputs for the two sub-networks, as a whole, the architecture produces the predictions and continues learning through training.

\textit{\textbf{Positional Encoding:}}
It includes the sequence order of the input in the embedding values by adding positional values. The positional encoding values also share the same dimension \textit{d} as the input embedding. The encodings are added using sine and cosine functions of different frequencies. The equations \ref{equation1} and \ref{equation2} show one of the many approaches to embed positional encoding, as mentioned by Vaswani et al.~\cite{vaswani2017attention}.
\vspace{-2mm}
\begin{equation}
\label{equation1}
    PE_{(pos,2i)} = \sin({pos}/{10000^{{2i}/{d}}})
\end{equation}

\vspace{-5mm}

\begin{equation}\label{equation2}
    PE_{(pos,2i+1)} = \cos({pos}/{10000^{{2i}/{d}}})
\end{equation}
where \textit{pos} is the position of the element in the sequence and \textit{i} is the \textit{th} dimension. 
\vspace{-1.5mm}
\subsubsection{\textbf{Encoder Network}}
\label{Encoder Network}
The encoder network consists of two sublayers. The first is the Self-Attention layer where the input dimensions are divided into \textit{N} number of identical sublayers, thus assembled to create a Multi-head Self-Attention mechanism described below. The second sublayer is a position-wise fully connected feed-forward network.

\begin{figure}[!t]
    \centering
    \includegraphics[width=0.8\linewidth]
    {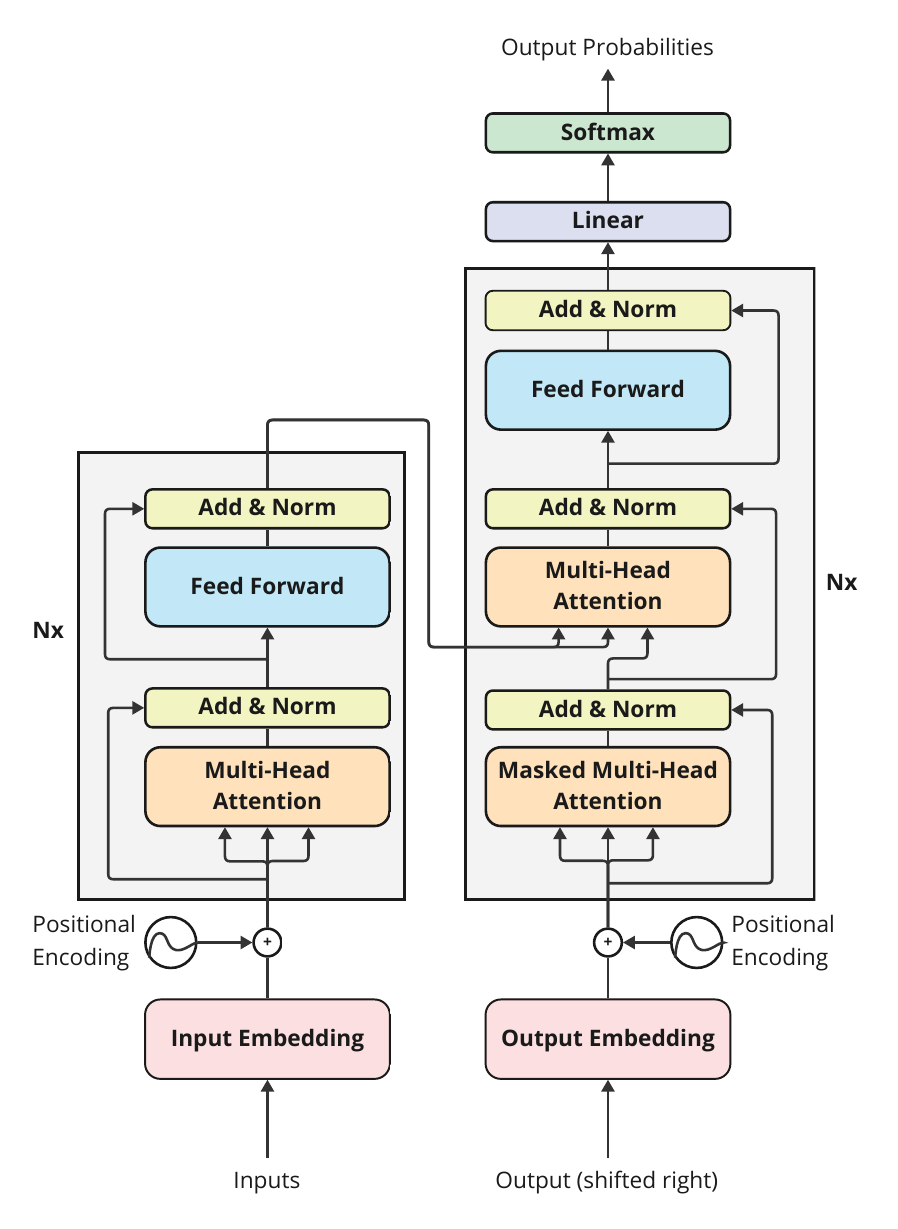}
    \caption{Transformer architecture as proposed by Vaswani et al. \cite{vaswani2017attention}}
    \label{fig:transformer}
     \vspace{-3mm}
\end{figure}

\textit{\textbf{Self-Attention Mechanism:}}
Attention values are the weights of each input entity concerning other input entities. They are quantified as the percentage of attention provided while encoding all the input sequences. The attention function is the mapping of \textit{Key(K), Query(Q), and Value(V)} where $K$, $Q$, and $V$ originate from the identical input sequences after positional encoding as shown 
in Fig.~\ref{fig:transformer}. The dimensions of both $K$ and $Q$ are $d_k$, and the dimension of $V$ is $d_v$. The setup allows each position in the input sequence to attend all positions in the same sequence to dynamically weigh and integrate information across the entire sequence, as shown in the steps below~\cite{han2022survey}.
\begin{itemize}
    \item \textit{Step 1:} Calculate the similarity between Query and Key vectors using the dot product $Q.K^T$; \textit{$K^T$ here is the transposition of the matrix K.}
    \item \textit{Step 2:} Normalize the score to maintain the stability in the gradient by dividing the similarity score as $\frac{Q.K^T}{\sqrt{d_k}}$
    \item \textit{Step 3:} Translate the scores into probabilities using the softmax as: $softmax(\frac{Q.K^T}{\sqrt{d_k}})$
    \item \textit{Step 4:} Obtain the Value matrix by multiplying $V$ with the probabilities, which gives the attention weight of each input entity as given in equation \ref{equation3}: 
\end{itemize}
 \vspace{-2mm}
\begin{equation}
\label{equation3}
   Attention(Q,K,V) = softmax{(\frac{QK^T}{\sqrt{d_k}})}V\ 
\end{equation}
 
\textit{\textbf{Multihead Attention Mechanism:}}
The three identical input values as $Q$, $K$, $V$ are multiplied with their respective weights. The resultant matrices are divided into multiple chunks (basically, the $d\_model$ is broken into smaller chunks), and self-attention values are calculated. They are called heads (\textit{h}) and are concatenated and multiplied with an attention weight to form multi-head attention values.

\textit{\textbf{Residual Connection and Normalization:}}
In both encoder and decoder networks, the layer normalization is followed by residual connections around each of the two sublayers in the encoder and around each of the three sublayers in the decoder. The residual connection, which is also called skip connection, allows the input to a particular layer to be added to its output by skipping a layer in between. This helps to alleviate the vanishing gradient problem, facilitate deeper networks, and improve learning capability. 

The Layer Normalization technique~\cite{ba2016layer} in DL is used to improve the training speed and help the network learn more effectively by improving internal covariate shifts (shifts in the distribution of activations as data moves through the network during training). With normalization, for each training sample, the mean and variance are computed across all neurons in a layer. The output from residual connection and normalization is given as equation \ref{equation4}:  
\vspace{-2mm}
\begin{equation}
\label{equation4}
    LayerNorm (X + Sublayer(X))
\end{equation}
where $X$ is the input sequence after positional encoding and $Sublayer(X)$ is the function of the layer preceding the residual connection (e.g., Attention layer, FFN layer, etc.) 

\textit{\textbf{Position-wise Feed Forward Network (FFN):}}
In the encoder and decoder network, there is one sublayer of FFN, which operates independently on each position to capture local and positional non-linear relationships. It consists of two linear transformations and one non-linear activation function, denoted as the following function below as equation \ref{equation5}. 
\vspace{-3mm}
\begin{equation}
\label{equation5}
    FFN(\textbf{X})= \sigma(W_1X+b_1)W_2+b_2
\end{equation}
where $X$ is the input to the FFN layer, $W_1$ and $W_2$ are the two linear transformations and $b_1$ and $b_2$ are the corresponding biases and $\sigma$ is the non-linear activation function like RELU \cite{agarap2018deep}, GELU \cite{hendrycks2016gaussian}, etc.
\vspace{-1.5mm}
\subsubsection{\textbf{Decoder Network}}
\label{Decoder Network}
The decoder network maintains the same structural stack as the encoder but incorporates two distinct modifications. The first one is the added sublayer called Masked Multi-Head Self-Attention, which takes input elements (mentioned as output in Fig.~\ref{fig:transformer}), which are shifted right, and future tokens are masked (this is because each token can only be influenced by previous tokens but not future ones). The second one is the Multi-Head Self-Attention mechanism implemented in the sublayer, followed by the masked self-attention sublayer, where the key and value pairs are passed from the output of the encoder network. This allows every position in the decoder to extract the attention of all positions in the input sequences at the encoder network.

\textit{\textbf{Linear, Softmax and Output:}}
The output of the whole network is a linear vector of predictions of the next element produced after processing the input and output sequences, along with attention values and FFN processing. Using the Softmax function in the transformer architecture helps determine the percentage for considering each entity. Thus, it is used to normalize the vector numbers into probabilities and ensure that the output is a valid probability distribution summing up to one.

\vspace{-2.5mm}
\subsection{\textbf{Pre-Trained Transformers}}
Pre-trained transformers are advanced ML models initially trained into a large corpus of data to learn the general representations. The concept of pre-trained models came into existence because DL models with many parameters need a much larger dataset to fully train the parameters. However, creating large-scale labeled datasets is a great challenge and costly. On the other hand, large-scale unlabeled datasets are relatively easier to create. Therefore to leverage this, pre-training approaches are typically applied to unlabeled data, known as self-supervised learning, which enables models to first learn the general representation from a large volume of data and use the learned representation to perform domain-specific downstream tasks by fine-tuning it. Various pre-trained models have been developed in recent years. Here, we discuss only the models that are used in the malware analysis domain. 

\textit{\textbf{Bidirectional Encoder Representation from Transformer (BERT):}}
BERT is a groundbreaking pre-trained encoder-only transformer model introduced by Devlin et al.~\cite{kenton2019bert}, which revolutionized the NLP domain. It can understand the context of a word based on all its surroundings (both in left and right directions) in all model layers. Fine-tuning BERT adapts BERT to specific downstream tasks by continuing the training process on a task-specific dataset, which is a part of the transfer learning approach. Based on various aspects like size, speed, efficiency, and data requirements, there are a lot of variants of BERT, such as DistilBERT~\cite{sanh2019distilbert}, 
RoBERTa~\cite{liu2019roberta}, AlBERT~\cite{lan2019albert}, etc. 

\textit{\textbf{CANINE:}}
CANINE~\cite{clark2022canine} is a pre-trained transformer-based tokenization-free neural encoder trained on character-level sequences for language representation. Unlike other pre-trained models trained on tokenization approaches like words, sentences, or other forms of tokenizations, CANINE is pre-trained using character sequences without explicit tokenization or vocabulary. There are two pre-trained CANINE models - CANINE-s (pre-trained using subword loss) and CANINE-c (pre-trained on characters with autoregressive character loss).

\textit{\textbf{Vision Transformer (ViT):}}
\label{vision transformer}
ViT~\cite{dosovitskiy2020image} applies the transformer architecture based encoder directly to sequences of image patches, treating each patch as a token. It introduced a new way to handle 2D images by flattening and projecting them into an embedding space like word embedding in text processing. 

\textit{\textbf{Generative Pre-trained Transformer (GPT2):}}
GPT-2~\cite{radford2019language} significantly expanded the original transformer architecture by adding many parameters and training data. It uses a stacked decoder-only architecture to generate text by predicting the next word. As GPT-2 was trained with extensive and diverse data from Internet, it is able to generate more coherent and contextual text across multiple domains.

\vspace{-2.5mm}

\section{Systematization: transformers in malware Analysis}
\label{Systematization: transformers in malware Analysis}
In this section, we introduce taxonomies based on the comprehensive application of transformers, the use of transformers to solve challenges, variations of transformers applied to malware analysis, and the diverse feature representations and correlations processed with the transformers. 


In Fig.~\ref{fig:malwaretasks-taxonomy} we show the different types of malware analysis tasks performed using transformers and in  Fig.~\ref{fig:transformer-taxonomy} a detailed taxonomic structure diagram. 
The taxonomy diagram consolidates and illustrates various modules and custom architectural enhancements used to represent different feature correlations. The visual structure maps out how diverse transformer architectures, from standard to custom-enhanced models, are tailored to capture and process a wide array of intricate patterns and correlations across different applications. We also present the extended in-depth scope of this topic in Appendix section 6 where we discuss in detail about how different transformers are adapted, enhanced and fine tuned for malware anaylsis tasks, along with evaluation and practical implementation insights.

The rest of this section is organized as follows.
In Subsection~\ref{Use Cases of transformers} we discuss the application of transformer models to perform various malware analysis-based end tasks like detection, classification, etc. 
In Subsection~\ref{TransformerAddressingChallenges} we discuss how the application of transformers is incorporated to address certain challenges in malware analysis. Then, in Subsection~\ref{transformers Used in Modeling} we categorize different types of transformers applied to address specific tasks in malware analysis, and in Subsection~\ref{Feature Representation} we discuss and categorize feature input types and representation techniques. Finally, in Subsection~\ref{Datasets Inventory} we present the dataset inventory. 

\vspace{-1.5mm}
\subsection{\textbf{Applications of Transformers in Malware Analysis}}
\label{Use Cases of transformers}

\begin{figure}
    \centering
    \includegraphics[width=1.0\linewidth]
    {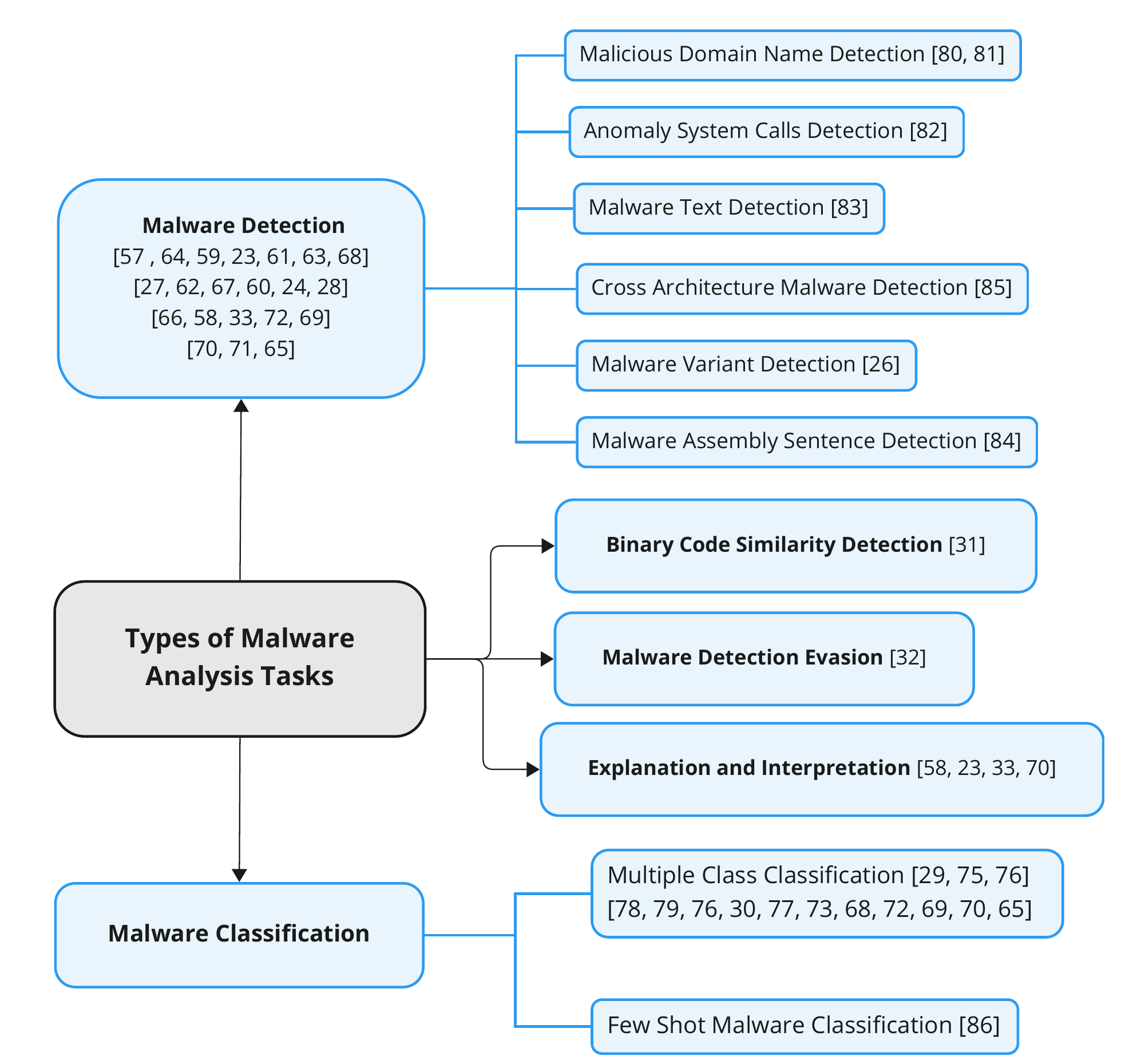}
    \vspace{-4mm}
    \caption{Types of Malware Analysis Tasks performed using Transformers}
    \label{fig:malwaretasks-taxonomy}
     \vspace{-5mm}
\end{figure}

Our analysis shows that the adoption of transformer architectures, within the malware analysis domain, supports and enhances the conventional security tasks (see 
Fig.~\ref{fig:malwaretasks-taxonomy}). From the figure, we can notice that a significant portion of the research has focused on malware detection ~\cite{chen2020android,hu2020exploit,long2021detecting,li2021mad,csahin2021malware,dehunting,wangwang2021network,bellante2021victory,fan2021heterogeneous,ullah2022explainable,qi2022mdfa,ghourabi2022security,seneviratne2022self,barut2022r1dit,pandya2023malware,deng2023transmalde,li2023iot,saracino2023graph,ravi2023vit4mal,jo2023malware,piadatrans,trizna2023nebula} as an end goal while harnessing the transformers. These approaches vary significantly employing different methodologies, feature representation techniques, and model combinations which are discussed in further sections. Additionally, some of these detection tasks have been extended not only to perform classification~\cite{seneviratne2022self,barut2022r1dit,saracino2023graph, jo2023malware,trizna2023nebula,moon2021directional,or2021pay,rahali2021malbert,li2022efficient,chen2022malicious,demirkiran2022ensemble,park2022vision,rahali2023malbertv2,belal2023global} but also to provide explanation and interpretation~\cite{hu2020exploit,li2021mad,ullah2022explainable,jo2023malware}, thereby enhancing the transformer's application to security analysis. Beyond malware-specific detection objectives, we uncovered a range of other detection-focused approaches in the field of malware analysis. These include malicious domain name detection\cite{yang2022n} \cite{gogoi2023dga}, anomaly system call detection~\cite{guan2021malware}, malware-related text detection from threat reports~\cite{shahid2020devising}, 
malware variant detection~\cite{lu2022research}, 
malware assembly sentence detection~\cite{demirci2022static}, and cross-architecture based malware detection in IoT 
systems~\cite{hamad2021bertdeep}. Moreover, a few unique approaches have addressed more complex aspects such as evasion 
techniques~\cite{hu2021single} and 
binary code similarity detection~\cite{li2023gental}, further demonstrating the broad spectrum of applications for transformer technology in addressing a variety of malware threats. 

Building upon the diverse end goals achieved through transformer architecture, most researchers have primarily utilized transformers for \textbf{generating contextual representations or embeddings} for input sequences (see Section \ref{GeneratingRobustEmbeddin} for additional discussion). This method utilizes the transformer's proficiency in capturing both local and long-range dependencies, crucial for processing complex data formats effectively. In addition to generating embeddings, numerous approaches have been proposed focusing on other innovative applications of transformers. 
For instance, Hu et al.~\cite{hu2021single} developed MalGPT, fine-tuned on benign files, to generate benign-looking perturbations that could evade detection systems. This approach represents a strategic shift towards using transformers not just for detection, but for enhancing evasion techniques. 
Similarly, Shahid et al.~\cite{shahid2020devising} leveraged the transfer learning approach to extract critical cybersecurity terms from malware reports, thereby improving automated forensic analysis. Furthermore, Demirkıran et al.~\cite{demirkiran2022ensemble} introduced an ensemble of pre-trained transformers, BERT and CANINE (applied CANINE for the first time in the malware analysis domain), configured as a bagging-based Random Transformer Forest (RTF) to showcase the applicability of transformer forest which could outperform state-of-the-art models (performing multi-class classification) even with highly imbalanced datasets. 

 The transformative applications of transformers within the realm of malware analysis are profound and diverse which demonstrates their evolving role in advancing Cybersecurity defenses. 
 

\vspace{-2.0mm}
\begin{tcolorbox}[colframe=gray!50!black, colback=gray!10, fonttitle=\bfseries, sharp corners, width = \linewidth, boxrule=0.5mm, boxsep=0mm, left=2mm, right=2mm]
\small\textit{\textbf{Takeaway: }Transformers' versatility across multiple analysis tasks—ranging from detection and classification to evasion and interpretation—demonstrates their evolving role in enhancing cybersecurity defenses.}
\end{tcolorbox}
\vspace{-3mm}
\vspace{-2mm}
\subsection{\textbf{Transformers Addressing Challenges in Malware Analysis: Efficiency, Effectiveness, and Adaptability}} 
\label{TransformerAddressingChallenges} 

To tackle the challenge of emerging malware variants and various obfuscation techniques, Lu et al.~\cite{lu2022research} developed a comprehensive approach for creating malware variant datasets. With that new dataset, they perform adversarial training of the BERT-based detection model to enhance the robustness and generalization abilities to work under obfuscation conditions. They disassembled samples into the assembly language and inserted API call sequences while maintaining the original code execution structure. This obfuscation technique involves inserting code and then extracting API so that the obfuscated malware can be used as a variant of the original malware. This technique helped to create an adversarial data set and retrain the model to make it more aware and robust against obfuscation techniques. 

To address the processing efficiency challenges posed by the transformer processing requirements, Li et al.~\cite{li2023iot} improved the training time efficiency and detection accuracy of the transformer-based malware detector by improving the standard transformer architecture. They introduced an additive attention mechanism - an improved version of the self-attention mechanism. Unlike the standard encoder, instead of calculating the dot product among Query($Q$), Key($K$), and Value($V$), which has a computational complexity of $O(N^2)$, they introduced a linear method to capture the context. Initially, the context information in $Q$ is \textbf{compressed} and \textbf{summarized} into a global query vector $q$ and it interacts with the matrix $K$ to produce the global matrix $k$. The resultant global context holding $k$ is multiplied by the value matrix $V$ to produce the attention values. In addition, they also replaced the standard residual connection and layer normalization with residual weight parameters $\alpha_i$. This adjustment also alleviates the gradient problem by allowing the model to dynamically adjust the residual contribution during training. These modifications collectively reduce the computational complexity by 5 times and accelerate the convergence of the network.  

Furthermore, Ravi et al.~\cite{ravi2023vit4mal} addressed the limitation of adaptability of transformers in resource-constraint environments such as IoT. They developed a lightweight, resource-efficient Vision Transformer (ViT) based model, ViT4Mal, for IoT malware detection, optimizing it for edge devices by simplifying its architecture. The limitation was addressed by reducing the computational complexity and memory requirements of the model. They reduced the dimensionality of image patch vectors through a linear projection in the patch embedding layer, used a learnable position embedding to the patch embedding, streamlined fine-tuned transformer encoder blocks, and simplified the decoder network without using an extra class token, unlike the original ViT architecture. After deploying the proposed lightweight model on the FPGA board (applying specialized hardware optimizations as well), they were able to speed up the processing by 41 times while maintaining the detection accuracy as compared to the original ViT. 

To evaluate the cost-effectiveness of a transformer-based model for varying sequence lengths, Or-Meir et al.~\cite{or2021pay} compared it with various architectures, including LSTM, LSTM with Attention, Bi-LSTM and Bi-LSTM with Attention. They evaluated each model's time-cost effectiveness to accurately classify unseen malware with 22 varying input sequence lengths (ranging from 10 to 4000). Based on their observation, they found that the transformer's performance was costly and less effective for lower sequence lengths, whereas it performed exceptionally better with the increasing sequence length. This pattern highlights the transformer's suitability for handling larger data sequences.

Bu et al.~\cite{bu2023triplet} tackled the challenge of enabling deep neural networks to cope with an increasing number of unknown malware samples and the reliance on large labeled datasets for new malware. To address this, they developed a few-shot malware classification solution using a graph transformer trained with triplet loss~\cite{bu2020monte}. This approach strategically applies transformers to handle novel malware samples in a few-shot scenarios, which are likely to be in real-world scenarios. Their method involves generating a control flow graph (CFG) from the malware's assembly code. The graph transformer selectively weights the correlations between nodes to capture the functional characteristics of the malware. The triplet loss function forces the network to learn a disentangled representation by minimizing the distance between similar malware samples and maximizing the distance between dissimilar ones. This approach enhances the transformer's ability to distinguish between malware types with minimal training data, improving the robustness of malware detection systems against obfuscation and variability in malware signatures.

\vspace{-2mm}
\begin{tcolorbox}[colframe=gray!50!black, colback=gray!10, fonttitle=\bfseries, sharp corners, width = \linewidth, boxrule=0.5mm, boxsep=0mm, left=2mm, right=2mm]
\small\textit{\textbf{Takeaway: }Transformers are potentially capable of overcoming key challenges concerning computational efficiency, adaptation 
to obfuscation techniques, and resource usage reduction.}
\end{tcolorbox}
\vspace{-2mm}
\subsection{\textbf{Transformer Variants Used in Malware Analysis}}
\label{transformers Used in Modeling}
Transformer architectures have been widely customized to tailor them to specific malware analysis tasks. We have developed the taxonomy, shown in Fig.~\ref{fig:transformer-taxonomy} to systematically organize all the resulting variations and their applications in capturing various intricate patterns and correlations. We have also consolidated information about the varieties of pre-trained and custom-enhanced transformers that were applied to malware analysis in
Table~\ref{Table:pretrainedtransformers}.

\subsubsection{\textbf{Standard (Vanilla) Transformer Architecture}}
Our analysis shows that the use of modular components is more ubiquitous than the use of the transformer architecture as a whole. As we did not find the use of a decoder-only module, we discuss only the attention mechanism and encoder-only modules in the section below.

\textbf{\textit{Attention Mechanism:}}
This mechanism mathematically quantifies the importance of different segments of the input data and allows the model to learn intricate patterns and relationships within the sequences. In malware analysis, this allows the network to prioritize critical elements of input sequences, and dynamically adjust weights based on their relevance.
A key aspect of the application of attention to malware analysis is the incorporation of positional encoding, which is vital for understanding the order and context of operations within the code. Positional encoding adds the necessary context to the model, helping it recognize the importance of sequence positioning in detecting malicious patterns. Thus, numerous approaches leverage the capabilities of attention mechanism, to learn intricate relationships within the input sequences and analyze them to make the decisions~\cite{chen2020android, long2021detecting, or2021pay, wangwang2021network, qi2022mdfa, deng2023transmalde}.
In the context of transformer architectures, positional encoding typically uses 1D values to maintain the order of data sequences. However, to analyze network traffic bytes, 
Barut et al.~\cite{barut2022r1dit} innovated within this framework by employing both 1D positional encoding and 2D convolutional feature embedding in their Residual 1-D Image Transformer (R1DIT) model. The term 1-D Image in R1DIT refers to the conceptual treatment of traffic byte sequences as one-dimensional data arrays that are processed using methods associated with image data using convolutional neural networks. Thus, the 1D positional encoding extracts the temporal sequence, ensuring that the model recognizes the order in which network events occur whereas the 2D convolutional feature embedding extracts and learns spatial features. This dual approach significantly improves the model’s capability to detect malware based on network traffic behaviors.

\textbf{\textit{Encoder Only:}}
The encoder-only module consists of an attention mechanism and the position-wise feed-forward network sublayer. This setup is crucial in emphasizing significant input parts and capturing complex non-linear relationships, making it suitable for malware analysis. It leverages the attention mechanism to emphasize significant parts of the input and feed-forward network to capture local and positional non-linear relationships. While it is consistently used across various studies, its application is uniquely tailored to address specific challenges in malware analysis. For example, Oliveira et al.~\cite{dehunting} utilize the encoder's ability to enhance malware detection through advanced feature representation, while Li et al.~\cite{li2023gental} focus on decoding assembly instructions' syntactical and semantic nuances. Hu et al.~\cite{hu2020exploit} leverage the encoder to efficiently analyze IoT malware by harnessing self-attention from the encoder, and Trizna et al.~\cite{trizna2023nebula} utilize the encoder for processing dynamic behavior patterns in varying sequence lengths. Similarly, Li et al.~\cite{li2022efficient} and 
Guan et al.~\cite{guan2021malware} highlight the encoder’s role in understanding complex patterns in API call data and system call sequences, respectively. Moon et al.~\cite{moon2021directional} emphasize the encoder's proficiency in managing long sequences, using its self-attention mechanism to capture the complex dependencies and structural nuances within Control Flow Graphs (CFGs). These varied applications demonstrate the encoder’s versatility in advancing malware analysis.

\vspace{-1mm}
\begin{tcolorbox}[colframe=gray!50!black, colback=gray!10, fonttitle=\bfseries, sharp corners, width = \linewidth, boxrule=0.5mm, boxsep=0mm, left=2mm, right=2mm]
\small\textit{\textbf{Takeaway: }The attention mechanism and encoder-only modular components are widely utilized in malware analysis. These components play a crucial role in capturing long-term dependencies and relationships within malware data, making them effective for various malware tasks. }
\end{tcolorbox}

\begin{table*}[ht!]
\centering
\caption{Application of Pre-trained Transformers and Custom Enhancements/Improvements on Transformer Architecture for Malware Analysis}

\label{Table:pretrainedtransformers}
\vspace{-3mm}
\begin{adjustbox}{width=\textwidth}
\begin{tabular}{|m{1.2cm}m{3cm}m{3.5cm}m{8cm}m{8cm}|}
\hline
\rowcolor{gray!30}
\textbf{Date} & \textbf{Transformer Type} & \textbf{Transformer Module} & \textbf{Objective of the Study} & \textbf{Objective of Integrating Transformer} \\
\hline
\rowcolor{blue!10}
\multicolumn{5}{|c|}{\textbf{Pre-Trained Transformers}}\\
\hline
2020 & BERT~\cite{shahid2020devising} & Encoder Only & Forensic Analysis of Threat Reports & Semantic Extraction of Malicious terms \\

\rowcolor{gray!10}
2021 & BERT~\cite{bellante2021victory} & Encoder Only & Detect IoT Malware with Limited Training (Explore BERT Model) & Enhance Detection with Robust Embeddings \\

2021 & BERT~\cite{rahali2021malbert} & Encoder Only & Classify Android Malware & Enhance Classification with Robust Embeddings \\

\rowcolor{gray!10}
2021 & Fine Tuned BERT~\cite{hamad2021bertdeep} & Encoder Only & Cross-architecture IoT malware Detection (Explore Fine-Tuning of BERT Model) & Enhance Detection with Robust Embeddings \\

2021 & GPT-2~\cite{csahin2021malware} & Decoder Only & Explore GPT2 for Malware Detection & Enhance Detection with Robust Embeddings \\

\rowcolor{gray!10}
2021 & GPT-2~\cite{hu2021single} & Decoder Only & Malware Detection Evasion using Benign Looking Perturbations & Construct Adversarial Malware Examples in a Single shot Black-box Setting \\

2021 & BERT~\cite{ghourabi2022security} & Encoder Only & Detect IoT Network Intrusion (Explore BERT Model) & Enhance Detection with Robust Embeddings \\

\rowcolor{gray!10}
2021 & BERT~\cite{ullah2022explainable} & Encoder Only & Interpret IoT Malware Detection & Enhance Detection by Integrating with Multi-modal System \\

2022 &Fine Tuned BERT  ~\cite{lu2022research} & Encoder Only & Construct Malware variants and Robustify Malware Detection & Enhance Robustness against Obfuscated Techniques \\

\rowcolor{gray!10}
2022 &  Ensemble of BERT and CANINE  ~\cite{demirkiran2022ensemble} & Encoder Only & Explore Random Transformer Forest for Highly Imbalanced Class Dataset for Malware Classification & Enhance Classification with Robust Embeddings \\

2022 & GPT-2  ~\cite{demirci2022static} & Decoder Only & Explore GPT-2 as Natural Langugage Model to Detect Malicious Assembly Sentences & Enhance Detection by Integrating with Multi-modal System \\

\rowcolor{gray!10}
2022 & ViT  ~\cite{chen2022malicious} & Encoder + Lambda Attention & Malware Classification with Reduced Computational Complexity and Memory usage using Lambda Attention & Enhance Malware Classification with Robust Embedding \\

2022 & ViT~\cite{park2022vision} & Encoder Only + Patch Encoding & Malware Classification with Enhanced Existing Model using Patch Encoding & Enhance Malware Classification with Robust Embedding \\

\rowcolor{gray!10}
2023 & ViT ~\cite{seneviratne2022self} & Encoder Only & Explore Self-Supervised Learning and Transfer Learning for Downstream Tasks of Malware Classification & Enhance Malware Classification with Robust Embedding \\

2022 & BERT  ~\cite{rahali2023malbertv2} & Encoder Only & Classify Android Malware & Enhance Classification with Robust Embeddings \\

\rowcolor{gray!10}
2022 & BERT  ~\cite{saracino2023graph} & Encoder Only & Detect and Classify Android Malware & Enhance Detection and Classification of Android Malware with Robust Embeddings \\

2023 & BERT, DistilBERT, RoBERTa, ALBERT~\cite{pandya2023malware} & Encoder Only & Comparison of BERT based Models to Detect Malware & Enhance Detection with Robust Embeddings \\

\rowcolor{gray!10}
2023 & CANINE-c ~\cite{gogoi2023dga} & Encoder Only & Explore CANINE Model in Malicious Domain Name Detection & Enhance Malicious Domain Detection with Robust Embedding \\

2023 & ViT ~\cite{jo2023malware} & Encoder Only + Attention & Interpret Malware Detection and Classification using Attention Maps & Enhance Malware Classification with Robust Embedding \\
\hline
\rowcolor{blue!10}
\multicolumn{5}{|c|}{\textbf{Custom Enhancement and Improvements}} \\
\hline
\rowcolor{gray!10}
2021 & Novel Transformer ~\cite{li2021mad} & Galaxy Transformer & Explore Unique Arrangement of Star Transformers to Enhance Malware Detection and Interpretation against Obfuscations Techniques& Complex Sequence Processing and Generate Robust Embedding \\

2023 & Novel Transformer~\cite{fan2021heterogeneous} & Heterogeneous Temporal Graph Transformer (HTGT) & Explore Combination of Transformers to Model Malware Evolution and Propagation Pattern & Enhance Malware Detection with Robust Embedding \\

\rowcolor{gray!10}
2023 & Novel Transformer~\cite{bu2023triplet} & Graph Transformer & Enhance Few-shot Classification with Less Volume of Data & Enhance Effectiveness of Few-shot Malware Classification with Robust Embeddings \\

2023 &  Improved Vanilla Transformer ~\cite{li2023iot} & Additive Attention + Residual weight parameters & Explore Custom Improvements to Enhance Detection & Increase Efficiency of Model by Reducing Computational Complexity \\

\rowcolor{gray!10}
2023 & Improved Vanilla Transformer ~\cite{piadatrans} & AdaTrans & Integrate Inter-component Communication (ICC) to Enhance Detection & Enhance Malware Detection with Robust Embedding \\

2023 & Improved ViT Transformer~\cite{belal2023global} & B\_ViT & Explore Global-local Attention to Enhance Malware Classification and Resilience to Polymorphic Obfuscation & Enhance Classification by Integrating with Multi-modal system \\

\rowcolor{gray!10}
2023 & Improved ViT Transformer~\cite{ravi2023vit4mal} & Lightweight ViT : VT4Mal & Explore Adaptability in Resource-constrained Environment by Developing Lightweight Vision Transformer & Enhance Detection with Robust Embedding and Explore Adaptability of Transformer in Low-Resource Environment\\
\hline
\end{tabular}
\end{adjustbox}
\vspace{-4mm}
\end{table*}

\subsubsection{\textbf{Pre-Trained Transformers}}
\label{pretrained transformers}
In malware analysis, there is a broad use of pre-trained models as shown in Table~\ref{Table:pretrainedtransformers}, focusing on \textit{encoder-only} and \textit{decoder-only} architectures, alongside a mix of various attention mechanisms and embeddings. Here are the different types of pre-trained transformers used in the malware field.

\textbf{\textit{BERT and BERT Variants: }}
BERT is used to generate robust embeddings by providing more nuanced indications of malware text~\cite{shahid2020devising}, focusing on deep contextual characteristics from an NLP perspective to detect malware~\cite{rahali2021malbert, rahali2023malbertv2, saracino2023graph}, addressing the challenge of detecting IoT malware with limited training data~\cite{bellante2021victory}, and processing contextual understanding to analyze network traffic~\cite{ullah2022explainable, ghourabi2022security}. Pandya et al.~\cite{pandya2023malware} compared the embeddings generated from BERT, DistillBERT, RoBERTa, and AlBERT using opcode sequences to classify malware classes, demonstrating superior accuracy over context-free embeddings. 

\textbf{\textit{Fine-Tuned BERT: }}BERT has also been fine-tuned with specific data, like opcode sequences extracted from executable files~\cite{hamad2021bertdeep} and by constructing obfuscated and unobfuscated malware variants~\cite{lu2022research}, which reflect the real-world malware behavior post-obfuscation, to perform the downstream tasks of malware detection and classification.

\textbf{\textit{CANINE: }}The character-level processing ability of CANINE-c is leveraged by Gogoi et al.~\cite{gogoi2023dga} to effectively analyze malicious domain names without tokenization. Demirkıran et al.~\cite{demirkiran2022ensemble} experimented on an ensemble of BERT and CANINE leveraging BERT's deep contextualized representations and CANINE's flexibility in handling raw text data without tokenization and also addressed the issue of imbalanced malware datasets for malware classification.

\textbf{\textit{GPT-2: }}GPT's advanced natural language processing capabilities with its attention mechanism have been leveraged to understand the context and relationship within opcode sequences~\cite{csahin2021malware} for malware detection. Assembly instructions extracted from \texttt{.text} sections of PE files are treated as sentences and documents to label as benign or malicious~\cite{demirci2022static} using GPT-2. In contrast, Hu et al.~\cite{hu2021single} used GPT-2 to generate adversarial malware examples in a single-shot black-box setting by adding \textit{benign-looking} perturbations in malwares.

\textbf{\textit{Vision Transformer (ViT) and Variants: }}
There are innovative approaches that utilize the abilities of Vision transformer (ViT)~\cite{dosovitskiy2020image} to capture long-range dependencies across the entire image without being limited by the local receptive fields as that in Convolutional Neural Networks (CNNs). 

Seneviratne et al.~\cite{seneviratne2022self} employed a self-supervised approach using a ViT-based Masked Auto Encoder, SHERLOCK, where the input image patches are masked and the model is trained to reconstruct the masked patches. Once the self-supervised training of SHERLOCK is completed, the learned representations are then transferred to fine-tune 3 different downstream models performing binary classification, multi-class (47 categories), and family classification (696 categories). The model achieved high accuracy in all 3 classification tasks outperforming models like ResNet, DenseNet, and MobileNetV2. In this study, the transfer learning approach using learned representations for downstream tasks appears faster, more convenient, and less data-heavy as compared to training those models from scratch.  

Chen et al.~\cite{chen2022malicious} introduced Lambda attention in the Encoder, which reduced the computational complexity from quadratic to linear by abstracting the interaction between query, key, and value into lambda functions and effectively aggregating the context from the input sequence. Thus, they simplified the complexity from $O(N^2)$ to $O(N)$. Park et al.~\cite{park2022vision} also introduced a technique distinct from the standard ViT. By adding patch by embedding along with the encoder which splits the input into patches and applies a transformer encoder directly, the approach adds an extra encoding step for each patch, which enriches the input data with more focused local features and positional information. 

Furthermore, Jo et al.~\cite{jo2023malware} used ViT to enhance malware detection and interpretability. The input images are passed through ViT to interpret the attention-based focus on different parts of the images and find the most relevant features for detection using the attention map from the ViT model.  

\vspace{-1mm}
\begin{tcolorbox}[colframe=gray!50!black, colback=gray!10, fonttitle=\bfseries, sharp corners, width = \linewidth, boxrule=0.5mm, boxsep=0mm, left=2mm, right=2mm]
\small\textit{\textbf{Takeaway: }The integration of pre-trained models provides a powerful foundation for malware detection workflows, reducing the need for large-scale labeled datasets while at the same time improving accuracy with fine-tuning, making them highly practical for real-world applications.}
\end{tcolorbox}

\subsubsection{\textbf{Custom Enhancements to Standard Architecture}}
Many researchers have focused on enhancing the existing transformer architecture by introducing novel architectural changes and unique arrangements as shown in Table~\ref{Table:pretrainedtransformers}. Li et al.~\cite{li2021mad} proposed a novel Galaxy transformer based on the arrangements of star 
transformers~\cite{guo2019star}  at the fundamental level, inspired from the heavenly bodies, creating three components based on the organization of star transformers: satellite-planet (to understand the basic blocks), planet-star (to understand assembly functions), and star-galaxy transformer (to understand the comprehensive semantics). 

Bu et al.~\cite{bu2023triplet} proposed a novel graph transformer designed to handle control flow graphs. It incorporates a directional embedding mechanism to capture sequential nature of control flow within the graph. It also adapts multi-head attention to prioritize critical attack paths and functional characters of malware samples. It integrates triplet loss function to learn a disentangled representation which ensures similar malware samples are embedded closely. With these integrations, they propose few-shot based malware classifier that mitigates the dependency on sample volume and performs distance-based malware classification. 

In Android malware analysis, Fan et al.~\cite{fan2021heterogeneous} proposed a novel Heterogeneous Temporal Graph transformer (HTGT), a combination of novel heterogeneous spatial transformer (to capture heterogeneity attention over each node and edge) and novel heterogeneous temporal transformer (to aggregate its historical sequences of a given node attentively), specifically designed to model malware's evolution and propagation pattern to learn their latent representation. HTGT (or called Dr. Droid) outperformed multiple state-of-the-art comparable models and showcases more than 98\% true positive rate in detecting novel malware samples. The model was deployed in the anti-malware industry to serve over 700 million mobile users worldwide against evolving android malware attacks.

Besides custom architectures and unique arrangements, there are several improvements made to the existing Vanilla transformer as well. In IoT malware analysis, to address the issue of computational complexity, 
Li et al.~\cite{li2023iot} introduced additive attention, which converts the quadratic complexity into a linear, as discussed in Section~\ref{TransformerAddressingChallenges}. The authors claim that their approach decreases the complexity five times as compared to the standard architecture. Similarly, Pi et al.~\cite{piadatrans} proposed AdaTrans, an adaptive transformer based on adaptive multi-head attention network to detect malware that uses inter-component communication. Adaptive attention modifies how attention scores are computed by adjusting the weights based on relevance offering a more flexible way to focus only on the most informative parts.
\begin{table*}[!t]
\centering
\caption{A Comprehension of Feature Representation Techniques Utilizing Various Transformer Models for Diverse Analytical Objectives}
\vspace{-3mm}
\label{Table:TaxTable}
\begin{adjustbox}{width=\textwidth}

\begin{tabular}{
|r|c|
>{\columncolor{gray!10}}c >{\columncolor{white}}c >{\columncolor{gray!10}}c >{\columncolor{white}}c >{\columncolor{gray!10}}c >{\columncolor{white}}c >{\columncolor{gray!10}}c >{\columncolor{white}}c >{\columncolor{gray!10}}c >{\columncolor{white}}c >{\columncolor{gray!10}}c|
>{\columncolor{white}}c >{\columncolor{gray!10}}c >{\columncolor{white}}c|
>{\columncolor{gray!10}}c >{\columncolor{white}}c >{\columncolor{gray!10}}c >{\columncolor{white}}c >{\columncolor{gray!10}}c >{\columncolor{white}}c >{\columncolor{gray!10}}c|
>{\columncolor{white}}c >{\columncolor{gray!10}}c >{\columncolor{white}}c >{\columncolor{gray!10}}c >{\columncolor{white}}c >{\columncolor{gray!10}}c >{\columncolor{white}}c >{\columncolor{gray!10}}c >{\columncolor{white}}c|
>{\columncolor{gray!10}}c >{\columncolor{white}}c >{\columncolor{gray!10}}c >{\columncolor{white}}c|
>{\columncolor{gray!10}}c >{\columncolor{white}}c >{\columncolor{gray!10}}c|l|
}
\hline
\multirow{10}{*}{\textbf{Surveyed Works}}
& \multirow{10}{*}{\begin{tabular}[c]
{@{}c@{}}\textbf{Year}\end{tabular}} 
& \multicolumn{11}{c|}{\begin{tabular}[c]{@{}c@{}}\textbf{Feature Types}\end{tabular}} 
& \multicolumn{3}{c|}{\begin{tabular}[c]{@{}c@{}}\textbf{Input Types}\end{tabular}} 
& \multicolumn{7}{c|}{\begin{tabular}[c]{@{}c@{}}\textbf{Feature Co-relations Types}\end{tabular}}  
& \multicolumn{9}{c|}{\begin{tabular}[c]{@{}c@{}}\textbf{Transformer Modules}\end{tabular}} 
& \multicolumn{4}{c|}{\begin{tabular}[c]{@{}c@{}}\textbf{Analysis Objectives}\end{tabular}} 
& \multicolumn{3}{c|}{\begin{tabular}[c]{@{}c@{}}\textbf{Analysis Methods}\end{tabular}} 
& \multirow{10}{*}{\begin{tabular}[c]{@{}c@{}}\textbf{Transformer}\end{tabular}} \\
\cline{3-39}

&       

&{\rotatebox[origin=l]{90}{1. API \& System Call}}
& \rotatebox[origin=l]{90}{2. Assembly code}
& \rotatebox[origin=l]{90}{3. Binary Sequence}
& \rotatebox[origin=l]{90}{4. Domain name}
& \rotatebox[origin=l]{90}{5. File System Info}
& \rotatebox[origin=l]{90}{6. Function call}
& \rotatebox[origin=l]{90}{7. Manifest File}
& \rotatebox[origin=l]{90}{8. Network Behavior}
& \rotatebox[origin=l]{90}{9. Opcode sequence}
& \rotatebox[origin=l]{90}{10. System Behavior}
& \rotatebox[origin=l]{90}{11. Others}

& {\rotatebox[origin=l]{90}{1. Graph}}
& \rotatebox[origin=l]{90}{2. Image}
& \rotatebox[origin=l]{90}{3. Text}

& {\rotatebox[origin=l]{90}{1. Functional}}
& \rotatebox[origin=l]{90}{2. Semantic}
& \rotatebox[origin=l]{90}{3. Sequential}
& \rotatebox[origin=l]{90}{4. Spatial}
& \rotatebox[origin=l]{90}{5. Structural}
& \rotatebox[origin=l]{90}{6. Syntactic}
& \rotatebox[origin=l]{90}{7. Temporal}

&{\rotatebox[origin=l]{90}{1. Attention}}
& \rotatebox[origin=l]{90}{2. Encoder Only}
& \rotatebox[origin=l]{90}{3. Decoder Only}
& \rotatebox[origin=l]{90}{4. Graph Transformer}
& \rotatebox[origin=l]{90}{5. HST}
& \rotatebox[origin=l]{90}{6. Star Transformer}
& \rotatebox[origin=l]{90}{7. Temporal Transformer}
& \rotatebox[origin=l]{90}{8. Custom Attention}
& \rotatebox[origin=l]{90}{9. Custom Changes}

&{\rotatebox[origin=l]{90}{1. Detection}}
& \rotatebox[origin=l]{90}{2. Classification} 
& \rotatebox[origin=l]{90}{3. Explanation}  
& \rotatebox[origin=l]{90}{4. Others}  

& \rotatebox[origin=l]{90}{1. Static}
& \rotatebox[origin=l]{90}{2. Dynamic} 
& \rotatebox[origin=l]{90}{3. Network}  
&

\\ \hline
Chen et. al.~\cite{chen2020android} & 2020 &
&& && && && $\surd$& &&
&& $\surd$&
&$\surd$& && && &
$\surd$&& && && && &
&$\surd$& &&
$\surd$&& &
Vanilla Transformer \\

Shahid et. al.~\cite{shahid2020devising} & 2020 &
&& && && && & &$\surd$&
&& $\surd$&
&$\surd$& && && &
&$\surd$& && && && &
&& &$\surd$&
$\surd$&& &
BERT \\ 

Hu et. al.~\cite{hu2020exploit} & 2020 &
&& && && && $\surd$& &&
&& $\surd$&
&$\surd$& &&$\surd$ && &
&$\surd$& && && && &
$\surd$&& $\surd$&&
$\surd$&& &
Vanilla Transformer \\ \hline

Moon et. al.~\cite{moon2021directional} & 2021 &
&$\surd$& && && && & &&
$\surd$&& &
$\surd$&& &&$\surd$ && &
&$\surd$& && && && &
&$\surd$& &&
$\surd$&& &
Vanilla Transformer \\ 

Hamad et. al.~\cite{hamad2021bertdeep} & 2021 &
&& && && && $\surd$& &&
&& $\surd$&
&$\surd$& && && &
&$\surd$& && && && &
$\surd$&& &&
$\surd$&& &
Fine Tuned BERT\\ 

Long et. al.~\cite{long2021detecting} & 2021 &
$\surd$&& && && &$\surd$& & &&
&& $\surd$&
&& $\surd$&& && &
$\surd$&& && && && &
$\surd$&& &&
&$\surd$& &
Vanilla Transformer \\ 

Guan et. al.~\cite{guan2021malware} & 2021 &
$\surd$&& && && && & &&
&& $\surd$&
&& $\surd$&& && &
&$\surd$& && && && &
$\surd$&& &&
&$\surd$& &
Vanilla Transformer \\ 

Or-Meir et. al.~\cite{or2021pay} & 2021 &
$\surd$&& && && && & &&
&& $\surd$&
&& $\surd$&& && &
$\surd$&& && && && &
&$\surd$& &&
&$\surd$& &
Vanilla Transformer \\

Li et. al.~\cite{li2021mad} & 2021 &
&$\surd$& && && && & &$\surd$&
&& $\surd$&
$\surd$&$\surd$& &$\surd$&$\surd$ && &
&& && &$\surd$& && &
$\surd$&& $\surd$&&
$\surd$&$\surd$& &
Galaxy Transformer \\ 

Rahali et. al.~\cite{rahali2021malbert} & 2021 &
&& && && $\surd$&& & &&
&& $\surd$&
$\surd$&$\surd$& && $\surd$&& &
&$\surd$& && && && &
&$\surd$& &&
$\surd$&& &
BERT \\ 

Sahin et. al.~\cite{csahin2021malware} & 2021 &
&$\surd$& && && && & &&
&& $\surd$&
&$\surd$& && &$\surd$& &
&& $\surd$&& && && &
$\surd$&& &&
$\surd$&& &
GPT-2 \\ 

Hu et. al.~\cite{hu2021single} & 2021 &
&& && && && $\surd$& &&
&& $\surd$&
&$\surd$& && && &
&& $\surd$&& && && &
&& &$\surd$&
$\surd$&& &
GPT-2 \\ 

Oliveira et. al.~\cite{dehunting} & 2021 &
$\surd$&& && && && & &&
&& $\surd$&
&$\surd$& $\surd$&& &$\surd$& &
&$\surd$& && && && &
$\surd$&& &&
$\surd$&$\surd$& &
Vanilla Transformer \\ 

Wangwang et al.~\cite{wangwang2021network} & 2021 &
&& && && &$\surd$& & &&
&& $\surd$&
&$\surd$& && && &
$\surd$&& && && && &
$\surd$&& &&
&$\surd$& &
Vanilla Transformer \\ 

Bellante et. al.~\cite{bellante2021victory} & 2021 &
$\surd$&$\surd$& && $\surd$&& &$\surd$& & $\surd$&&
&& $\surd$&
&$\surd$& $\surd$&& && &
&$\surd$& && && && &
$\surd$&& &&
$\surd$&$\surd$& $\surd$&
BERT \\ 

Fan et al.~\cite{fan2021heterogeneous} & 2021 &
$\surd$&& && && $\surd$&& & &$\surd$&
$\surd$&& &
&& &$\surd$& && $\surd$&
&& && $\surd$&& $\surd$&& &
$\surd$&& &&
$\surd$&$\surd$& &
HTGT \\ \hline

Ullah et. al.~\cite{ullah2022explainable} & 2022 &
&& && && &$\surd$& & &&
&& $\surd$&
&$\surd$& && && &
&$\surd$& && && && &
$\surd$&& $\surd$&&
&& $\surd$&
BERT  \\ 

Li et. al.~\cite{li2022efficient} & 2022 &
$\surd$&& && && && & &&
&& $\surd$&
&& $\surd$&& && &
&$\surd$& && && && &
&$\surd$& &&
$\surd$&$\surd$& &
Vanilla Transformer \\

Chen et. al.~\cite{chen2022malicious} & 2022 &
&& && && && $\surd$& &&
&$\surd$& &
&& &$\surd$& && &
&$\surd$& && && &$\surd$& &
&$\surd$& &&
$\surd$&& &
ViT (Vision Transformer) \\ 

Lu et. al.~\cite{lu2022research} & 2022 &
$\surd$&& && && && & &&
&& $\surd$&
$\surd$&$\surd$& && && &
&$\surd$& && && && &
$\surd$&& &&
&$\surd$& &
Fine Tuned BERT\\

Qi et. al.~\cite{qi2022mdfa} & 2022 &
$\surd$&& && && && & &&
&& $\surd$&
&$\surd$& && && $\surd$&
$\surd$&& && && && &
$\surd$&& &&
&$\surd$& &
Vanilla Transformer \\ 

Demirci et. al.~\cite{demirci2022static} & 2022 &
&$\surd$& && && && & &&
&& $\surd$&
&$\surd$& && && &
&& $\surd$&& && && &
$\surd$&& &&
$\surd$&& &
GPT-2 \\ 

Ghourabi et. al.~\cite{ghourabi2022security} & 2022 &
&& && $\surd$&& &$\surd$& & $\surd$&&
&& $\surd$&
&$\surd$& && && &
&$\surd$& && && && &
$\surd$&& &&
$\surd$&& $\surd$&
BERT \\ 

Demirkiran et. al.~\cite{demirkiran2022ensemble} & 2022 &
$\surd$&& && && && & &&
&&$\surd$&
&$\surd$& && && &
&$\surd$& && && && &
&$\surd$& &&
$\surd$&$\surd$& &
Random Transformer Forest \\ 

Seneviratne et. al.~\cite{seneviratne2022self} & 2022 &
&& $\surd$&& && && & &&
&$\surd$& &
&$\surd$& && && &
&$\surd$& && && && &
$\surd$&$\surd$& &&
$\surd$&& &
ViT \\

Park et. al.~\cite{park2022vision} & 2022 &
&& $\surd$&& && && & &&
&$\surd$& &
&$\surd$& && $\surd$&& &
&$\surd$& && && && $\surd$&
&$\surd$& &&
$\surd$&& &
ViT \\ 

Yang et. al.~\cite{yang2022n} & 2022 &
&& &$\surd$& && && & &&
&& $\surd$&
&$\surd$& $\surd$&& && &
$\surd$&& && && && &
$\surd$&& &&
$\surd$&& &
Vanilla Transformer \\ 

Barut et. al.~\cite{barut2022r1dit} & 2022 &
&& && && &$\surd$& & &&
&& $\surd$&
&& &$\surd$& && $\surd$&
$\surd$&& && && && &
$\surd$&$\surd$& &&
&$\surd$& &
Vanilla Transformer \\ \hline

Rahali et. al.~\cite{rahali2023malbertv2} & 2023 &
&& && && $\surd$&& & &&
&& $\surd$&
&$\surd$& && && &
&$\surd$& && && && &
&$\surd$& &&
$\surd$&& &
BERT \\ 

Gogoi et. al.~\cite{gogoi2023dga} & 2023 &
&& &$\surd$& && && & &&
&& $\surd$&
&$\surd$& && && &
&$\surd$& && && && &
$\surd$&& &&
$\surd$&& &
CANINE Transformer \\ 

Li et. al.~\cite{li2023gental} & 2023 &
&$\surd$& && && && & &&
&& $\surd$&
&$\surd$& && && &
&$\surd$& && && && &
$\surd$&& &&
$\surd$&& &
Vanilla Transformer \\ 

Pandya et. al.~\cite{pandya2023malware} & 2023 &
&& && && && $\surd$& &&
&& $\surd$&
&$\surd$& && && &
&$\surd$& && && && &
$\surd$&& &&
$\surd$&& &
BERT, DistilBERT, RoBERTa, AlBERT \\ 

Deng et. al.~\cite{deng2023transmalde} & 2023 &
$\surd$&& && &$\surd$& && & &&
$\surd$&& &
&$\surd$& && && &
$\surd$&& && && && &
$\surd$&& &&
$\surd$&& &
Vanilla Transformer \\ 

Li et. al.~\cite{li2023iot} & 2023 &
$\surd$&& && && && $\surd$& &&
&& $\surd$&
&$\surd$& && && &
&& && && &$\surd$& $\surd$&
$\surd$&& &&
$\surd$&& &
Improved Vanilla Transformer \\ 

Belal et. al.~\cite{belal2023global} & 2023 &
&& $\surd$&& && && & &&
&$\surd$& &
&& &$\surd$& && &
&& && && &$\surd$& $\surd$&
&$\surd$& &&
&$\surd$& &
Butterfly ViT \\ 

Saracino et. al.~\cite{saracino2023graph} & 2023 &
$\surd$&& && && && & &&
&&$\surd$&
&$\surd$& && && &
&$\surd$& && && && &
$\surd$&$\surd$& &&
$\surd$&& &
BERT \\ 

Ravi et. al.~\cite{ravi2023vit4mal} & 2023 &
&& $\surd$&& && && & &&
&$\surd$& &
&& &$\surd$& && &
&$\surd$& && && && &
$\surd$&& &&
$\surd$&& &
Lightweight ViT \\ 

Jo et al.~\cite{jo2023malware} & 2023 &
&& $\surd$&& && && & &&
&$\surd$& &
&$\surd$& && $\surd$&& &
$\surd$&$\surd$& && && && &
$\surd$&$\surd$& $\surd$&&
$\surd$&& &
ViT \\ 

Pi et al.~\cite{piadatrans} & 2023 &
&& && &$\surd$& && & &&
$\surd$&& &
&$\surd$& && && &
&& && && &$\surd$& &
$\surd$&& &&
$\surd$&& &
Improved Vanilla Transformer \\ 

Trizna et al.~\cite{trizna2023nebula} & 2023 &
$\surd$&& && $\surd$&& &$\surd$& & $\surd$&&
&& $\surd$&
&$\surd$& && && &
&$\surd$& && && && &
$\surd$&$\surd$& &&
&$\surd$& &
Vanilla Transformer \\ 

Bu et al.~\cite{bu2023triplet} & 2023 &
&$\surd$& && && && & &&
$\surd$&& &
$\surd$&& $\surd$&& && &
&& &$\surd$& && && &
&$\surd$& &&
$\surd$&& &
Proposed Transformer \\ \hline

\end{tabular}
\end{adjustbox}
\vspace{-5mm}
\end{table*}

Furthermore, several approaches optimize the Vision transformer (ViT)~\cite{dosovitskiy2020image}. 
Belal et al.~\cite{belal2023global} proposed Butterfly\_ViT, which incorporates a global-local attention mechanism, arranging the multiple-attention mechanism in such a way that it partitions images into segments to capture detailed local features and uses an entire image to extract global details in contrast to the standard ViT, which focuses uniformly across the whole image. However, Butterfly is computationally expensive, but harnessing parallel processing and leveraging the local-global representation seems to be a good approach. Ravi et al.~\cite{ravi2023vit4mal}, to cope with the limited resource constraints in IoT systems, proposed a ViT4Mal (a lightweight vision transformer) for edge devices by improving several aspects of standard ViT. They reduced the dimension in the patch embedding layer by projecting image segments linearly, streamlined the encoder by adjusting the number and complexity of the block used, and optimized hardware through quantization of model weights and activation, loop pipe-lining, and array partitioning.

Conclusively, from a higher level perspective, we observed that overall performance measures highlight the exceptional accuracy of transformer models in malware detection, often achieving accuracy greater than 95\%. Custom architectures that blend transformers with task-specific components have shown significant improvements. While high accuracy is common, there is often a trade-off regarding computational cost. Complex models, although requiring more resources, yield superior accuracy, demonstrating that the selection of a model should balance accuracy with available computational resources. For a more comprehensive understanding of transformer adaptations, their effectiveness, and performance measures, refer to the detailed insights in Appendix Sections 6.1 and 6.2.
\vspace{-1mm}
\begin{tcolorbox}[colframe=gray!50!black, colback=gray!10, fonttitle=\bfseries, sharp corners, width = \linewidth, boxrule=0.5mm, boxsep=0mm, left=2mm, right=2mm]
\small\textit{\textbf{Takeaway: }Custom enhancements to transformer architectures provide a significant leap in their application to malware analysis, making them more effective, scalable, and adaptable.}
\end{tcolorbox}

\vspace{-2mm}
\vspace{-3mm}
\subsection{\textbf{Feature Representations using Transformers}}
\label{Feature Representation}
When applying AI techniques to malware analysis, proper feature representation is a critical design choice, as it directly impacts the model's ability to accurately analyze malicious software. Using appropriate feature representations, transformers can better understand and interpret complex malware behaviors.

This section focuses on different types of feature and correlation representations using transformer models, as outlined in Table~\ref{Table:TaxTable}, demonstrating their flexibility in analyzing malware through diverse combination of features. We first categorize the features into 11 types as listed below, examining how these are extracted, generated, and transformed. Next, we discuss the integration of transformers in multi-modal systems, highlighting how combining textual, visual, and structural data representations can enhance malware analysis. Then, we categorize correlations into 7 different representation types and discuss how different studies harness transformers to generate robust embeddings for the different input features.

\subsubsection{\textbf{Feature and Input Types}}
As shown in Table~\ref{Table:TaxTable}, we have categorized features into 11 feature types focusing on how they are used as the \textit{medium} of representation. We also analyze approaches for extracting, generating, or transforming features.

\textbf{\textit{Binary Sequences:}}
These are the sequence of 0s and 1s, representing the executable codes. In malware analysis, using 
streams of binaries
from the executables, the streams are transformed into images and patches of images~\cite{seneviratne2022self,park2022vision,belal2023global,ravi2023vit4mal,jo2023malware} to visualize unseen aspects of malware.

\textbf{\textit{Opcode Sequences:}}
Opcodes are the low-level instruction sets executed by the CPU (hexadecimal form). In malware analysis, these sequences allow one to identify patterns of malware by analyzing the contextual relationship between the sequence of opcodes. The sequences are employed in diverse ways: as n-gram words for textual analysis~\cite{li2023iot}, to represent code blocks~\cite{chen2020android}, converted into images for visual analysis~\cite{chen2022malicious}, mapped to functions for structural insights~\cite{hu2020exploit}, and utilized as text sequences~\cite{hu2021single, hamad2021bertdeep, pandya2023malware}.

\textbf{\textit{Assembly Codes:}}
Assembly codes are expressed in low-level programming languages, closely related to the machine code and are readable by humans. In malware analysis, such readable assembly codes are usually obtained from executable files 
using disassembling tools and used as a sequence of text~\cite{li2023gental, csahin2021malware, demirci2022static}. The goal of such transformation is to extract contextual relationships among the opcodes and operands in the codes. Assembly codes can also be combined with other features extracted from different dynamic analyses~\cite{li2021mad,bellante2021victory} to enrich the set of features. Moreover, Moon et al.~\cite{moon2021directional} used Control Flow Graphs (CFGs) generated from assembly codes to represent jumps and function calls. Bu et al.~\cite{bu2023triplet} extracted CFGs from assembly to simulate malware attack paths and propagation.

\textbf{\textit{API and System Call Sequences:}}
API calls are requests made by programs to external libraries or services, and the system calls are requests made directly to the operating system (OS), which also represent interactions with the hardware. Several approaches have used API call sequences~\cite{li2022efficient, lu2022research, qi2022mdfa,demirkiran2022ensemble} and system call sequences~\cite{guan2021malware, or2021pay, dehunting} that are usually extracted from sandbox (dynamic analysis) environments. API and system call sequences can also be combined with other 
features~\cite{bellante2021victory,li2023iot,trizna2023nebula,long2021detecting} to construct long call sequences and extract insights about the malware's actions and intentions.
On the other hand, Fan et al~\cite{fan2021heterogeneous} and Deng et al.~\cite{deng2023transmalde} proposed approaches to convert API call sequences into graphs as feature representations, and conversely, Saracino et al.~\cite{saracino2023graph} proposed an approach to generate API call sequences from the API call graphs. These contrasting approaches highlight different analytical perspectives. In contrast, generating sequences from graphs can simplify the analysis by linearizing the relationships, making it easier to apply sequence-based machine learning models.

\textbf{\textit{Function Calls:}}
To uncover hidden or latent malware behaviors, sub-graphs generated from critical function calls (API calls that indicate potentially malicious behaviors) are used in the approach by Deng et al.~\cite{deng2023transmalde}, whereas Pi et al.~\cite{piadatrans} integrated Inter-component communication in the sensitive function call sub-graphs.

\textbf{\textit{File System Information:}}
A few approaches use file system information and other features to enrich the features' strength. Relevant features of file system include a large variety of static information, such as file metadata (size, creation date, and format)~\cite{bellante2021victory}, file operation path~\cite{trizna2023nebula}, header information~\cite{ghourabi2022security}, and file properties such as name, size, and list of strings~\cite{ghourabi2022security}, as well as dynamic information, such as file operation type~\cite{trizna2023nebula}, imported and exported functions, and properties sections~\cite{ghourabi2022security}.

\textbf{\textit{Manifest File Information:}}
In Android systems, manifest files play an important role as they provide essential information about the OS, permissions the app requires, activities used, services used, broadcast receivers used, content providers used, version supported, etc. In malware analysis, Rahali et al.~\cite{rahali2021malbert, rahali2023malbertv2} proposed an approach using manifest files to reveal suspicious permissions or uncommon features, and Fan et al.~\cite{fan2021heterogeneous} proposed an approach using manifest files as one of the features to map the malware propagation and reveal attack path. 

\textbf{\textit{Network Behaviors:}}
Malware analysis may also analyze network-based patterns or actions related to how malware communicates or behaves over the network. Network properties used in malware analysis include protocol information, data in transfer, 
network addresses~\cite{bellante2021victory}, connection ports and server names~\cite{trizna2023nebula}, dynamic feature items (extracted with strace and tcdump from system calls and IP activities)~\cite{long2021detecting}, traffic data (IP, HTTP, DNS, unencrypted TLS record headers)~\cite{barut2022r1dit} and packet files (.pcap)~\cite{wangwang2021network, ullah2022explainable, ghourabi2022security}.  

\textbf{\textit{System Behaviors:}}
In malware analysis, the behavior of malware that showcases the interactions and activities on the OS level are tracked as system behaviors to detect malicious activities. System behaviors like processes (execution, termination, manipulation)~\cite{bellante2021victory}, file modification (creation, deletion, or alteration)~\cite{bellante2021victory}, registry key accesses, access type and 
key-value~\cite{bellante2021victory,trizna2023nebula}, monitoring logs and behaviors (sensor data, OS logs)~\cite{ghourabi2022security} are used to map the system behaviors. In malware analysis, these behaviors are used as one of the features rather than using them as a standalone feature set.

\textbf{\textit{Domain Names:}}
Domain names play a crucial role in network malware analysis as they can be used to identify command-and-control servers, phishing sites, and other malicious destinations. Features, such as domain names and n-gram location, are often used to classify the semantic patterns into benign and malicious categories. Gogoi et al.~\cite{gogoi2023dga} leverage character-level analysis using the CANINE transformer to analyze domain names without tokenization, enhancing the detection accuracy for malicious domains. Similarly, Yang et al.~\cite{yang2022n} utilize n-gram features to capture the contextual information within domain names, further improving the classification of malicious domains.

\textbf{\textit{Others:}}
In addition to the commonly used features, some approaches incorporate unique features for malware analysis. Static features, like printable string, PE imports, and PE header numerical, were used by Li et al.~\cite{li2021mad} to observe the behavior, purpose, and origin of a file. Fan et al.~\cite{fan2021heterogeneous} used Android-based dynamic features such as loaded dex files, connected URLs, generated texts, and the application's social information (application name, affiliation, market, signature) to map the malware propagation behaviors. Shahid et al.~\cite{shahid2020devising} used natural language features obtained from malware threat report sentences to automate the detection of malicious terms from the NLP reports.

\vspace{-1mm}
\begin{tcolorbox}[colframe=gray!50!black, colback=gray!10, fonttitle=\bfseries, sharp corners, width = \linewidth, boxrule=0.5mm, boxsep=0mm, left=2mm, right=2mm]
\small\textit{\textbf{Takeaway: }The ability of transformers to process a wide range of feature types enhances their capability to analyze malware from multiple angles, providing a holistic understanding of malicious behaviors. This flexibility allows for a combination of static, dynamic, and network-based analysis within a single framework.}
\end{tcolorbox}
\vspace{-4mm}
\subsubsection{\textbf{Transformers Integration in Multi-Modal System}}

Multi-modal ML refers to the integration and processing of data from multiple modalities, such as text, images, and graphs, to enhance the analysis and understanding of complex information. In the context of malware analysis, this approach is intriguing as it allows for a more comprehensive understanding of malware behavior by leveraging different types of data, which can reveal various aspects of malicious activities that might be missed when considering a single modality. In malware analysis, integrating transformers or modules of transformers with multi-modal data opens up new possibilities for exploring the synergistic potential of multi-modal combinations from multiple perspectives. 
 

Recent approaches for  \textit{windows malware} detection show that integrating transformers in multi-modal systems is effective. The MDFA framework proposed by 
Qi et al.~\cite{qi2022mdfa} leverages the Bi-LSTM and attention mechanisms to analyze API call sequences, capturing both temporal dependencies and critical sequence patterns. Demirci et al.~\cite{demirci2022static} enhanced static malware detection by combining stacked BiLSTM and GPT-2 models, treating assembly instructions(as word, sentence, and document) as textual data to extract syntactic and semantic features. The 
Belal et al.'s~\cite{belal2023global} Global-Local Attention-Based Butterfly Vision Transformer model employs global and local attention mechanisms to process image and textual data, enabling the detection of complex malware patterns through a comprehensive analysis and offering resilience to polymorphic obfuscation. 


In the \textit{IoT malware} analysis domain, 
Ullah et al.~\cite{ullah2022explainable} developed an explainable malware detection system that employs transformers-based transfer learning and multi-modal visual representation, combining BERT (with a self-developed FAST extractor) extracted textual features with malware-to-image conversions (using self-developed BRIEF descriptor) for visual analysis. Their approach, which also incorporates CNN for deep feature extraction, emphasizes the importance of explainable AI in making detection processes transparent and reliable. To detect \textit{android malware}, Oliveira et al.~\cite{dehunting} proposed a multi-modal deep learning network, called Chimera. The multi-network processes three different feature types - \textit{permission and intents}, \textit{API call sequences}, and \textit{apk images}, individually fed into Deep Neural Network (DNN), Transformer Encoder Network (TN), and Convolutional Neural Network (CNN), respectively. Further, at the fusion layer, the collected representations from each sub-networks are concatenated as an ensemble of those subnetworks thus outperforming the classical ML methods.


Additionally, to analyse \textit{linux malware}, Guan et al.~\cite{guan2021malware} proposed a hybrid LSTM-Transformer model for system call anomalies detection. This hybrid approach highlights the value of combining LSTM's capability to handle sequential data with transformers' strength in learning global dependencies. Furthermore, Barut et al.~\cite{barut2022r1dit} introduced the Residual 1-D Image Transformer (R1DIT) for privacy-preserving malware traffic classification. This model leverages raw data transformation and attention-based modules to classify different malware types and benign traffic without interfering with IP addresses, port numbers, and the payload. Their results demonstrate superior accuracy and generalization, especially for handling new traffic types like TLS 1.3.

\vspace{-4mm}
\begin{tcolorbox}[colframe=gray!50!black, colback=gray!10, fonttitle=\bfseries, sharp corners, width = \linewidth, boxrule=0.5mm, boxsep=0mm, left=2mm, right=2mm]
\small\textit{\textbf{Takeaway: }Transformers integrated into multi-modal systems offer a powerful approach to malware analysis, as they can draw on diverse data types to provide an holistic view of malicious behaviors, particularly in handling complex, polymorphic, or evasive malware.}
\end{tcolorbox}

  \begin{figure*}
        \centering
        \includegraphics[width=\textwidth]
        {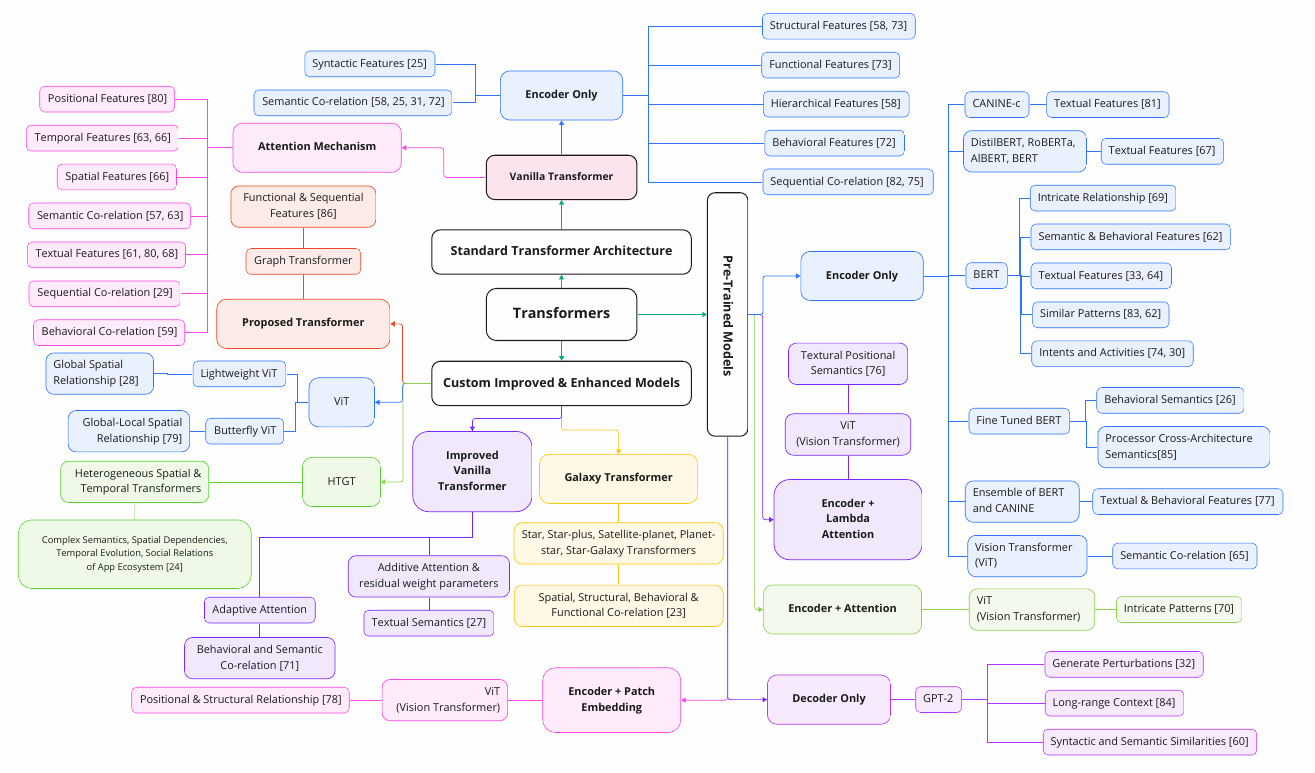}
        \caption{Taxonomy of transformers that are applied in malware analysis to capture various correlations and relationships from the input features }
        \label{fig:transformer-taxonomy}
        \vspace{-3mm}
    \end{figure*}

\subsubsection{\textbf{Transforming Feature Correlations Through Transformer Integration}}

Representing feature correlations involves capturing the relationships and dependencies between different data inputs. This is crucial because understanding these relationships helps in accurately analyzing malware behaviors. Transformers are particularly well-suited for this task due to their ability to model complex interactions within the data. As detailed in 
Table~\ref{Table:TaxTable}, we have categorized the correlations, studied in the malware domain, into seven types. This categorization provides a structured overview of how various studies address the representation of features correlations in their analyses. Additionally, Fig. \ref{fig:transformer-taxonomy} shows how different feature correlation representation techniques are captured with the integration of various transformer types.

\textbf{\textit{Functional Correlation:}} It refers to the relationship between different parts of the code that perform specific tasks or functions. In malware analysis, research has focused on functional relationship extraction by using assembly codes converted as control flow graphs~\cite{moon2021directional, bu2023triplet}, API and system call sequences~\cite{lu2022research}, manifest files~\cite{rahali2021malbert}, and basic blocks as hierarchical structure (organization of data or entities in layers of increasing complexity or specificity)~\cite{li2021mad} to map malware intents and activities. 

\textbf{\textit{Semantic Correlation:}} It refers to the meaningful relationship between data based on their meaning or context. Most of the reviewed literature~\cite{chen2020android, shahid2020devising, hu2020exploit, hamad2021bertdeep,
li2021mad,
rahali2021malbert,
csahin2021malware,
hu2021single,
dehunting,
wangwang2021network,
bellante2021victory,
ullah2022explainable,
lu2022research,
qi2022mdfa,
demirci2022static,
ghourabi2022security,
demirkiran2022ensemble,
seneviratne2022self,
park2022vision,
yang2022n,
rahali2023malbertv2,
gogoi2023dga,
li2023gental,
pandya2023malware,
deng2023transmalde,
li2023iot,
saracino2023graph,
jo2023malware,
piadatrans,
trizna2023nebula} extracts the semantic relationship using a variety of features. Semantic representations are widely used to observe malware behaviors, complex patterns, hierarchical structure, long-range dependencies, malware intents \& activities, etc.
\begin{table*}[!t]
\centering
\caption{Inventory of datasets that are used in malware analysis leveraging transformer architecture. \\
\textit{\scriptsize Labels mentioned as \(^1\) to \(^7\) are further explained in Section \ref{Datasets Inventory}. Columns descriptions: a) Dataset Name - Names of the datasets initially given by the creating author, b) Dataset Content Information: Brief information about the dataset content, c) Environment: In malware analysis, to which specific environment the datasets are used for the studies, d) End-goals: Main objectives as end-goals and their relevant citation of the literature.}}
\label{table:datasets}
\vspace{-3mm}
\begin{adjustbox}{width=\textwidth}
\begin{tabular}{|LMDR|}
\hline
\rowcolor{gray!5}
\textbf{Dataset Name} & \textbf{Dataset Contents Information} & \textbf{Environment} & \textbf{Malware Analysis End-Goal Task}  \\ \hline


Omnidroid~\cite{martin2019android}& 22,000 real malware and benign Android applications &Android & Malware Detection~\cite{dehunting}\\

\rowcolor{rowcolor}
Tencent Security Lab & Real-world data that contains 82,831 Android apps collected from 55 app stores/marketplaces/websites&Android &Malware Detection\cite{fan2021heterogeneous} \\

MalNet\cite{freitas2022malnet}&1,262,024 malware images extracted from real-world Android applications in AndroZoo &Android & Malware Detection \& Classification\cite{seneviratne2022self}\\


\rowcolor{rowcolor}
IoTPOT\cite{pa2015iotpot, pa2016iotpot}& 124,799 executables captured by IoTPOT honeypot & IoT& Malware Detection\cite{hu2020exploit, ravi2023vit4mal} \& Explanation\cite{hu2020exploit}\\


IoT-23\cite{garcia2020iot23}& 20 malicious and 3 benign traffic files created by AIC Lab by Avast Software &IoT &Malware Detection\cite{wangwang2021network} \\

\rowcolor{rowcolor}
Android Malware Genome\cite{malgenomeprojectAndroidMalware}\(^1\)& Android malware samples and associated metadata& IoT&Malware Detection\cite{bellante2021victory}\\

CICInvesAndMal2019\cite{taheri2019extensible}& & IoT& Malware Detection \& Explanation\cite{ullah2022explainable}\\

\rowcolor{rowcolor}
ECU-IoHT\cite{ahmed2021ecu}&  Traces of several types of attacks launched to target medical devices&IoT& Intrusion Detection\cite{ghourabi2022security} \\

ToN-IoT\cite{booij2021ton_iot}&  IoT network intrusion that combines information from pcap files, Bro logs, sensor data, and OS logs&IoT &Intrusion Detection\cite{ghourabi2022security} \\

\rowcolor{rowcolor}
Edge\_IIoTset\cite{ferrag2022edge}& Data from IoT devices such as temperature and humidity sensors, heart rate sensors, flame sensors, etc&IoT &Intrusion Detection\cite{ghourabi2022security} \\

EMBER\cite{anderson2018ember}&  Features extracted from 1.1M binary files distributed as malicious and benign& IoT&Intrusion Detection\cite{ghourabi2022security}\\

\rowcolor{rowcolor}
ICE\cite{fernandez2019intelligent}\(^2\)& Network analysis of set of ransomware attacks performed in an Integrated Clinical Environment (ICE)& IoT&Intrusion Detection\cite{ghourabi2022security}\\ 

    IMG\_DS\cite{kaggle2021iotmalware}& 14733 benign and 2486 malicious Android IoT image representation & IoT &Malware Detection\cite{ravi2023vit4mal} \\

\rowcolor{rowcolor}
Drebin\cite{arp2014drebin}& Malware and Benign Android apps collected from Android Malware Genome Project, Google Playstore, Chinese Markets, Russian Markets, Android websites, Malware forums and security blogs& Android, IoT&Malware Detection\cite{deng2023transmalde,saracino2023graph,piadatrans} \& Classification\cite{chen2020android,rahali2023malbertv2,saracino2023graph}\\

Androzoo\cite{allix2016androzoo} & 1046190 malware samples with 37 malware classes& Android, IoT& Malware Detection\cite{long2021detecting,dehunting,deng2023transmalde,jo2023malware}, Classification\cite{rahali2021malbert, rahali2023malbertv2,jo2023malware} \& Explanation\cite{jo2023malware}\\

\rowcolor{rowcolor}
CICMalDroid2020 or Maldroid\cite{mahdavifar2020dynamic}& 17,341 Android samples from VirusTotal, Contagio, AMD, and MalDozer & Android, IoT&Malware Detection\cite{ullah2022explainable,deng2023transmalde,saracino2023graph,piadatrans,jo2023malware}, Classification\cite{saracino2023graph,jo2023malware} \& Explanation\cite{ullah2022explainable,jo2023malware}\\

Malshare& Self Collected data that contains 38,427,440 basic blocks of 5 to 250 instructions on each& Windows&Malware Detection \& Explanation\cite{li2021mad} \\

\rowcolor{rowcolor}
Catak\cite{catak2019benchmark}& Windows API call sequences obtained within Cuckoo Sandbox and the corresponding malware classes of 7107 records&Windows &Malware Classification\cite{li2022efficient, demirkiran2022ensemble}\\

Sorel-20m\cite{harang2020sorel}& 672 benign and 822 malware samples & Windows & Malware Assembly Sentence Detection\cite{demirci2022static}\\

\rowcolor{rowcolor}
Oliveira\cite{de2023behavioral}\(^3\)& 42,797 malware and 1,079 benign API call sequences analysed dynamically&Windows &Malware Classification\cite{demirkiran2022ensemble} \\

MalImg\cite{nataraj2011malware}\(^4\)& MalImg by Vision Research Lab contains 9339 malware byteplot images & Windows& Malware Classification\cite{belal2023global}\\

\rowcolor{rowcolor}
BIG2015&BIG2015 by Microsoft contains 21,741 malware samples images &Windows &Malware Classification\cite{belal2023global} \\

Top-1000\cite{004e-v304-19}&Cotains over 47,000 portable executable both malware and benign imports &  Windows& Malware Classification\cite{belal2023global}\\

\rowcolor{rowcolor}
Speakeasy Dataset\cite{speakeasy2021}&Contains 93500 behavioral JSON format malware and benign data&Windows &Malware Detection \& Classification\cite{trizna2023nebula} \\

Avast-CTU Dataset\cite{bosansky2022avast}& Contains 400,000 samples in JSON format&Windows &Malware Detection \& Classification\cite{trizna2023nebula} \\

\rowcolor{rowcolor}
Malicious Code Dataset\cite{mcd2019}& Malicious Code Dataset (MCD) contains labeled 30,000 samples containing API sequences in XML format&Windows &Malware Detection \& Classification\cite{trizna2023nebula} \\

VirusTotal\cite{virustotal2021}& &Android, Windows &Malware Detection\cite{long2021detecting}, Classification\cite{or2021pay, demirkiran2022ensemble}, Detection Evasion\cite{hu2021single}\\

\rowcolor{rowcolor}
VirusShare\cite{virusshare}& PE format malicious code sample binaries&Android, Windows, IoT &Cross Architecture Malware Detection\cite{hamad2021bertdeep}, Malware Detection\cite{li2021mad,csahin2021malware,qi2022mdfa,deng2023transmalde,ravi2023vit4mal,piadatrans}, Malware Classification\cite{chen2022malicious,demirkiran2022ensemble,rahali2023malbertv2}, Explanation\cite{li2021mad}, Malware Variant Detection\cite{lu2022research}, Malware Assembly Sentence Detection\cite{demirci2022static}\\

Alexa\cite{ghodke2018alexa, ExpiredDomainsAlexaTop}\(^5\)&Alexa contains top 1 million legitimate domain names & Network&Malicious Domain Name Detection\cite{yang2022n, gogoi2023dga}\\

\rowcolor{rowcolor}
360 Network Security Lab Dataset\cite{dganetlab2022}\(^6\)&Contains malicious domain names publicly collected from different DGA families of malicious domain names& Network& Malicious Domain Name Detection\cite{yang2022n}\\

Stratosphere IPS\cite{stratosphere2015datasets}& Contains large public network capture data in .pacp format&Network&Malware Detection \& Classification\cite{barut2022r1dit}\\ 

\rowcolor{rowcolor}
CICIDS2017\cite{sharafaldin2018toward}& Contains raw capture data with the whole trace record throughout the day&Network &Malware Detection \& Classification\cite{barut2022r1dit}\\ 

Netlab Open Data Project\cite{netlabopendata2022}\(^7\)& Contains 1 million DGA  generated malicious domains along with the malware family which generated the domain name& Network& Malicious Domain Name Detection\cite{gogoi2023dga}\\

\rowcolor{rowcolor}
MalwareBazaar\cite{abuseMalwareBazaarMalware} & 86225 malware binary files&Linux, Windows &Binary Code Similarity Detection\cite{li2023gental}\\

Malpedia\cite{fraunhoferMalpediaFraunhofer}& 3158 malware binary files&Linux, Windows&Binary Code Similarity Detection\cite{li2023gental} \\

\rowcolor{rowcolor}
SecureNLP Challenge\cite{phandi2018semeval}&SubTask1, Semeval Task 8 from SecureNLP challenge dataset contains 11250 sentences from Advanced Persistent Threat (APT) reports & Network&Malware Texts Detection\cite{shahid2020devising} \\

Kaggle Microsoft Challenge\cite{ronen2018microsoft}& Kaggle Microsoft Malware Classification 
 Challenge contains 10868 assembly codes and binary codes (9 malware families)& Windows, Others& Malware Classification\cite{moon2021directional, chen2022malicious, park2022vision, belal2023global}, Few Shot Malware Classification\cite{bu2023triplet}\\

\rowcolor{rowcolor}
Self-Collected Data& 
Work \cite{hu2020exploit} - Malware samples from Honeypot and Benign Linux ARM samples, Work \cite{hamad2021bertdeep} - Previously used by HaddadPajouh et al.\cite{haddadpajouh2018deep}, Work \cite{guan2021malware} - Lab-generated dataset by Dymshits et al.\cite{dymshits2017process} (Stream of Vectors of integers of 300 length system call sequences),
Work \cite{li2021mad} - Benign executable collected from installation paths of software programs,
Works \cite{csahin2021malware, demirci2022static} - Benign files from Windows OS and Commando VM,
Work \cite{hu2021single, qi2022mdfa} - Benign files from Windows System files,
Works \cite{rahali2023malbertv2, deng2023transmalde, saracino2023graph} - Benign files collected from Google Play Store\cite{googleAndroidApps},
Work \cite{li2023gental} - Benign files collected from Software programs from Linux and Windows, Work \cite{pandya2023malware} - 2793 malware families with one or more samples per family \cite{Kim2018PEHeader}, Work \cite{li2023iot} - 30,000 API Calls, including 14,302 malicious samples and 15,698 benign samples. 9000 Opcodes samples of which 3500 samples are malicious and 5500 samples are benign & Android, Windows, IoT, Others& Malware Detection\cite{hu2020exploit,guan2021malware,li2021mad,csahin2021malware,qi2022mdfa,deng2023transmalde,saracino2023graph,li2023iot,pandya2023malware}, Malware Classification\cite{rahali2023malbertv2,saracino2023graph}, Explanation\cite{hu2020exploit,li2021mad}, Cross Architecture Malware Detection\cite{hamad2021bertdeep}, Malware Assembly Sentence Detection\cite{demirci2022static}, Malware Detection Evasion\cite{hu2021single}, Binary Code Similarity Detection\cite{li2023gental}\\ \hline

\end{tabular}
\end{adjustbox}
\vspace{-5mm}
\end{table*}

\textbf{\textit{Sequential Correlation:}} It refers to the relationships between the data points that depend on their order or sequence. In the malware domain, mainly the system and API call sequences~\cite{long2021detecting,
guan2021malware,
or2021pay,
dehunting,
bellante2021victory,
li2022efficient} as text along with the domain as n-gram words \cite{yang2022n} and the assembly codes as graphs \cite{bu2023triplet} are used to investigate the position-dependent malicious nature of malware.

\textbf{\textit{Spatial Correlation:}} It refers to the physical or logical relationship between data entities based on their spatial arrangement or proximity. In malware analysis, features are converted to images~\cite{chen2022malicious,
belal2023global,
ravi2023vit4mal} to observe spatial patterns, graphs~\cite{fan2021heterogeneous} are used to map dependencies, and texts~\cite{barut2022r1dit} are used to map the hierarchical structure \cite{li2021mad}. 

\textbf{\textit{Structural correlation:}} Unlike spatial correlation, structural correlation deals with the arrangement and connections between components, defining how they are put together to form a larger system. In the malware domain, there is the use of features as images~\cite{park2022vision, jo2023malware}, as 
graphs~\cite{moon2021directional}, features as texts to map the hierarchical structure~\cite{hu2020exploit, li2021mad} and malware intents \& activities~\cite{rahali2021malbert}.

\textbf{\textit{Syntactic Correlation:}} It is based on the structural rules i.e. syntax of data. In malware analysis, these correlations are observed in long text sequence analysis~\cite{csahin2021malware, dehunting}.

\textbf{\textit{Temporal Correlation:}} It refers to relationships between data entities that change over time. In the malware domain, capturing malware behaviors that evolve with 
time~\cite{fan2021heterogeneous}, 
tracking of the timing of API calls~\cite{qi2022mdfa}, and network behaviors~\cite{barut2022r1dit} using timestamps are experimented.

\vspace{-1mm}
\begin{tcolorbox}[colframe=gray!50!black, colback=gray!10, fonttitle=\bfseries, sharp corners, width = \linewidth, boxrule=0.5mm, boxsep=0mm, left=2mm, right=2mm]
\small\textit{\textbf{Takeaway: } The ability of transformers to capture a wide array of feature correlations enables them to detect more complex and sophisticated malware, as well as pave the way to model malware behaviors using multi-correlations. This makes them invaluable for defending against evolving security threats. }
\end{tcolorbox}

\vspace{-3mm}
\subsubsection{\textbf{Generation of Robust Embedding Representations using Transformers}}
\label{GeneratingRobustEmbeddin}
In this subsection, we cover approaches that use modules of standard or improved transformer architectures for robust embedding representations as high-dimensional data from different input types: images, graphs, and texts. These embeddings aim to capture critical information and relationships into a more manageable, higher-dimensional space.
\textit{\textbf{Images to Represent Features:}}\\
When processing images as input, the vision transformer 
(ViT)~\cite{dosovitskiy2020image} is most commonly used. As mentioned in Section~\ref{vision transformer}, the ViT is based on an encoder-only module. It processes patches from input images as tokens and generates the embedding representation in higher-dimensional space. The embeddings are further used as per the goal of the project. Utilizing ViT for feature representation from images is increasingly common. Such an approach, detailed in the \textit{Survey on Vision transformer} by Han et al.~\cite{han2022survey} and in the \textit{Survey of transformer in Vision} by Khan et al.~\cite{khan2022transformers}, i.e., representing features with ViT for robust embeddings are found in malware analysis as well. Though there are several approaches for  generating images to represent malware samples, the role of ViT usually appears once the input images are converted into patches and they are provided to ViT as input tokens to generate and represent the images into higher dimensional embeddings. Park et al.~\cite{park2022vision} used ViT to encode positional information of image patches and sequential information between the local features of the malware image. Unlike Park's classification objective, 
Seneviratne et al.'s~\cite{seneviratne2022self} goal is to learn malware image reconstruction and classify malware samples, but the core use of ViT is similar in both cases - to generate feature representations. In addition to the detection and classification task using the embeddings from ViT, Jo et al.~\cite{jo2023malware} propose to use a ViT attention map to provide explanations about the evaluation of malware samples. While such approaches use the standard ViT architecture, Chen et al.~\cite{chen2022malicious} proposed an improved ViT by introducing a Lambda layer. The Lambda layer in the ViT enhances its ability to learn positional relationships dynamically during training, unlike the original ViT which relies on static positional encodings. It also replaces the traditional self-attention mechanism with a more efficient linear function, reducing computational complexity and memory usage.\\
\textit{\textbf{Graphs to Represent Features:}}\\
In malware analysis, 
graphs are usually generated from different ways to map structural aspects of malware samples. Among the various approaches, the one by Saracino et al.~\cite{saracino2023graph} generates API call graphs from the Android APKs and converts them into API call sequences. Similarly, the approach by Deng et al.~\cite{deng2023transmalde} generates graphs from sensitive API and function calls to reflect latent behavioral patterns. The approach by Feng et al.~\cite{piadatrans} utilizes the features used in the approach by Deng et al.~\cite{deng2023transmalde} and also incorporates Inter-component communication (ICC) features to generate graphs based embeddings. Since API calls are often used to represent malicious patterns, in real-world instances, malware often exploits existing APIs in the system to mimic the normal behavior of benign apps to evade detection. So, by incorporating ICC patterns in addition to API call graphs, malware detection becomes more robust to model evasion due to the integral nature of component interaction within an application which is less susceptible to modification or removal by malicious intentions. Besides, the limitations of using only traditional CFG to generate embeddings are also highlighted by Moon et al.~\cite{moon2021directional}. They proposed a new method to incorporate edge information to preserve opcode semantics and add functional characteristics to the existing structural representations from CFG. This is achieved through a method that transforms opcodes into nodes. In addition, they utilize the DeepWalk algorithm~\cite{perozzi2014deepwalk} to traverse the CFG and create sequences that reflect the comprehensive relationships within the graph, including the newly added edge information. The sequences generated by DeepWalk are then embedded using a transformer encoder model, handling long sequences and capturing complex dependencies between nodes more effectively than traditional LSTM-based methods.\\
\textit{\textbf{Texts to Represent Features:}}\\
Like in NLP, in malware analysis, the transformer models are applied to sequences of textual features for detecting and classifying malicious behaviors. The approaches mentioned in this section primarily use transformers to focus on generating robust representation embeddings using different types of textual features. Using opcode sequences generated from disassembly, several approaches have been proposed~\cite{chen2020android, hu2020exploit, pandya2023malware, hamad2021bertdeep, csahin2021malware} that extract syntactic and semantic co-relation represented embeddings. 
Similarly, approaches have been proposed \cite{rahali2023malbertv2,trizna2023nebula,li2022efficient} that use API call sequences along with Manifest Files, Permissions, Services, and Intents~\cite{rahali2023malbertv2}, network connection data and registry access information~\cite{trizna2023nebula} to generate robust embeddings. To detect malicious domain names, the approach by Yang et al.~\cite{yang2022n} uses n-gram location and text information, and the one by Gogoi et al.~\cite{gogoi2023dga} that uses domain names. The approach by Ghourabi et al~\cite{ghourabi2022security} combines file information, header information, imported and exported functions, .pcap network files, sensor data, and OS logs to generate embeddings in the IoT domain. 
The approach by Bellante et al.~\cite{bellante2021victory} uses file metadata, binary code, network behavior, system behavior, and API call sequences to generate embeddings and also addresses the problem of fewer samples by using BERT in IoT malware analysis. Li et al.~\cite{li2023gental} introduced GenTAL, a generative denoising skip-gram (aimed to predict context words given a target word)  transformer~\cite{mikolov2013efficient} for binary code similarity detection by learning compact and meaningful representations from assembly codes.

\vspace{-1mm}
\begin{tcolorbox}[colframe=gray!50!black, colback=gray!10, fonttitle=\bfseries, sharp corners, width = \linewidth, boxrule=0.5mm, boxsep=0mm, left=2mm, right=2mm]
\small\textit{\textbf{Takeaway: } Transformer models are effectively used to generate robust embeddings from diverse input types—images, text, and graphs—in malware analysis. This enables transformers to capture complex relationships, enhancing malware detection and classification across multiple data modalities.}
\end{tcolorbox}
\vspace{-3.5mm}

\vspace{-2mm}
\subsection{\textbf{Datasets Inventory}}
\label{Datasets Inventory}
We tried our best to access and verify all the datasets that are included in the Table~\ref{table:datasets}. Android malware Genome (marked\(^{1}\) in Table \ref{table:datasets}) dataset is available but the efforts to update it is stopped as mentioned in the source of the dataset. The reference provided for the ICE Dataset (marked\(^{2}\)), Oliviera Dataset (marked\(^{3}\)), and MalImg Dataset (marked\(^{4}\)) are the references of work where the datasets are first used or introduced. For Alexa Dataset (marked\(^{5}\)), the provided link did not work, so we found another link and added it to the reference. Furthermore, for 360 Network Security Lab Dataset (marked\(^{6}\)) and Netlab Open Data Project Dataset (marked\(^{7}\)), the provided links were unreachable when we tried to access them. Besides the access concerns, we observed a few datasets that are widely used, such as the Drebin, Androzoo, Maldroid, VirusTotal, VirusShare, and Kaggle Microsoft Challenge datasets. In addition, researchers have created variations of the data sets according to their requirements for the experiments.

\section{Challenges and Future Prospects}
\label{sec:future}
Based on our analysis of the state of the art, we identify and categorize the limitations and future research directions into five broad categories, discussed below:

\textbf{Model and Architecture Limitations: }
Studies such as~\cite{shahid2020devising,hamad2021bertdeep, park2022vision, barut2022r1dit} present initial analyses of transformer architecture and attention mechanisms. As observed, with the increase in input sequence size (e.g., high-resolution images), the multi-head attention mechanism performance decreases, hindering the overall performance, so fewer multi-heads are used. Thus, a relevant research direction is the design of techniques for optimizing the multi-head attention to improve processing and inference time for specific tasks. 

\textbf{Inadequate Data and Feature Representation: }
The problem of imbalanced datasets and the need for diverse malware samples~\cite{moon2021directional, chen2020android, shahid2020devising, gogoi2023dga, demirci2022static, bu2023triplet} is also a critical issue in the field of malware analysis especially given the rise of sophisticated malware types. Therefore, the generation of synthetic malware exhibiting unseen characteristics would be critical. Work by Lu et al. focuses on the construction of malware variants by inserting API calls into the assembly code~\cite{lu2022research}. However, more work is needed focusing on generating synthetic malware samples using Generative AI, GANs, Auto-encoders, etc., opting for various obfuscation techniques, unseen characteristics, and intents of malware. 

\textbf{Multi-modality, Cross-Attention and RL: }
In malware analysis, most of the approaches use single feature types with inadequate pre-processing techniques
\cite{moon2021directional, hamad2021bertdeep, long2021detecting, guan2021malware, li2022efficient, ravi2023vit4mal, deng2023transmalde, piadatrans, saracino2023graph, seneviratne2022self, or2021pay, hu2021single}. Addressing such a challenge requires advanced feature engineering efforts. However, as an initial step toward such a direction, we observed a good focus on feature engineering, such as efficiently reducing the dimension of opcodes by converting them into decimal 
representation~\cite{chen2020android}, preserving edge/opcode information (which is usually lost)~\cite{moon2021directional}, slicing system call sequences into various lengths of subsequences to experiment with effective sequential processing
\cite{or2021pay}, and using multiple feature sets~\cite{li2021mad, ghourabi2022security}. 

Besides the existing efforts, there is still scope for future research on combining efficient and novel feature engineering techniques for multi-perspective representation of input samples. One of the directions is the design of multi-feature representation based multi-modal transformer network for malware analysis. 
Oliveira et al.~\cite{dehunting} experimented with multiple features like text and images, with multi-modal Deep Neural Network, CNN, and Transformer. However, they do not provide a combined perspective, in that they use an ensemble method at the fusion layer, which does not combine the multiple perspectives from the input. It rather uses a voting mechanism for the detection decision. There are fusion techniques, like cross-attention, fused attention, etc., investigated in other domains like vision~\cite{nagrani2021attention, zhang2023cross}, which can be extended to the field of malware analysis. This can help explore multiple perspectives using multi-feature representations (such as generating graph, image and text data from the same input sample or different samples) using multi-modal transformer networks.   

Also, experiments combining transformer attention mechanism and reinforcement learning as suggested by Ullah et al.~\cite{ullah2022explainable} for more robust malware detection can be an interesting avenue to explore. In malware analysis, the state representations can be a sequence of actions taken by the malware which could be features like visual, textual or graphs sequences extracted from feature types like API calls, file system changes, network activities, registry modifications, etc. The definition of actions could involve identifying malicious behavior, flagging suspicious flags, etc. On top of it, the reward function could be designed as the incentives when RL agent correctly identifies malicious intents. 
 Studies in other domains~\cite{chen2021decision, chen2022transdreamer} have shown that this integration through memory-based reasoning and  sequential behavior modeling, is effective for scalability and efficiency, which could be leveraged for malware analysis.   

\textbf{Robustness and Security: }
The vulnerabilities of DL models have been widely investigated because of their popularity. Similarly, there are investigations that demonstrate threats and vulnerabilities in transformer-based models. Some of the attacks are as follows: BAE ~\cite{garg2020bae}, a black box attack to generate adversarial examples using contextual perturbations from BERT; GBDA~\cite{guo2021gradient}, the first general purpose framework for gradient-based white-box attacks against text transformers; CWBA~\cite{liu2022character}, a character-level white-box adversarial attack; Vision Transformer based attack~\cite{wei2022towards}, a dual attack framework, which contains a Pay No Attention (PNA) attack and a PatchOut attack, to improve the transferability of adversarial samples across different ViTs; Positional Encoding based attack~\cite{gao2024pe}, an adversarial attack that manipulates the model by providing it with incorrect positional information enabling an evasion attack. These attacks demonstrate the inherent vulnerabilities of transformer-based models regardless of their application, which also impacts the malware analysis domain.

In malware analysis, the vulnerabilities of transformers based on adversarial attacks have not been well analyzed; however, the need for robustness has been highlighted~\cite{li2021mad, wangwang2021network, trizna2023nebula}. 
So, future research should focus on adversarial training, defensive mechanisms, evaluation of models' resilience against sophisticated attacks. In addition, the potential for malware evasion using obfuscation techniques, as well as its evolving nature~\cite{jo2023malware, hu2020exploit,fan2021heterogeneous} highlight the need for models that can effectively detect obfuscation techniques. Research is needed to design models capable of understanding complex obfuscation patterns.

\textbf{Deployment and Real-world Application: }
Since transformer-based models are computationally intensive, there are challenges related to the deployment of these models in practical settings, particularly on-edge devices~\cite{wangwang2021network, ghourabi2022security, belal2023global, ravi2023vit4mal}. Thus,  an important direction is the optimization of transformer models for various deployment environments like cloud, edge computing, and mobile platforms. Also, creating APIs that leverage transformer models for easy integration into security systems and tools, allowing for real-time malware detection and response in applications, is an interesting avenue for future research. 
\vspace{-3mm}
\section{Conclusions}
\label{sec:summary}

In this paper, we provide a comprehensive systematization of knowledge (SoK) on the use of transformers in malware analysis. 
Our analysis shows that transformers excel in capturing intricate patterns such as spatial, temporal, structural correlations, etc. across high-dimensional data, making them well-suited for detecting and classifying malware, analyzing binary code similarity, understanding evasion techniques etc. We also discuss the challenges associated with the application of transformers in malware analysis and present an inventory of datasets used in the domain to facilitate future research and development.


\section*{Acknowledgement}
This work is partially supported by the NSF grants 2230609, 2230610, 2416990 and 2229876.


\bibliographystyle{IEEEtran}

\begin{thebibliography}{100}
\providecommand{\url}[1]{#1}
\csname url@samestyle\endcsname
\providecommand{\newblock}{\relax}
\providecommand{\bibinfo}[2]{#2}
\providecommand{\BIBentrySTDinterwordspacing}{\spaceskip=0pt\relax}
\providecommand{\BIBentryALTinterwordstretchfactor}{4}
\providecommand{\BIBentryALTinterwordspacing}{\spaceskip=\fontdimen2\font plus
\BIBentryALTinterwordstretchfactor\fontdimen3\font minus \fontdimen4\font\relax}
\providecommand{\BIBforeignlanguage}[2]{{%
\expandafter\ifx\csname l@#1\endcsname\relax
\typeout{** WARNING: IEEEtran.bst: No hyphenation pattern has been}%
\typeout{** loaded for the language `#1'. Using the pattern for}%
\typeout{** the default language instead.}%
\else
\language=\csname l@#1\endcsname
\fi
#2}}
\providecommand{\BIBdecl}{\relax}
\BIBdecl

\bibitem{avtestmalwareStats}
{AV-TEST}, ``{Malware Statistics},'' \url{https://www.av-test.org/en/statistics/malware/}, [Accessed on Feb 15, 2024].

\bibitem{jang2014survey}
J.~Jang-Jaccard and S.~Nepal, ``{A survey of emerging threats in cybersecurity},'' \emph{Journal of computer and system sciences}, 2014.

\bibitem{aryal2021survey}
K.~Aryal, M.~Gupta, and M.~Abdelsalam, ``{A survey on adversarial attacks for malware analysis},'' \emph{arXiv preprint}, 2021.

\bibitem{gopinath2023comprehensive}
M.~Gopinath and S.~C. Sethuraman, ``{A comprehensive survey on deep learning based malware detection techniques},'' \emph{Computer Science Review}, 2023.

\bibitem{raff2018malware}
E.~Raff and et~al., ``{Malware detection by eating a whole exe},'' in \emph{Workshops at the AAAI conference on artificial intelligence}, 2018.

\bibitem{abijah2020vision}
S.~Abijah~Roseline and et~al., ``{Vision-based malware detection and classification using lightweight deep learning paradigm},'' in \emph{Computer Vision and Image Processing}.\hskip 1em plus 0.5em minus 0.4em\relax Springer, 2020.

\bibitem{mimura2022applying}
M.~Mimura and R.~Ito, ``{Applying NLP techniques to malware detection in a practical environment},'' \emph{International Journal of Information Security}, 2022.

\bibitem{tran2017nlp}
T.~K. Tran and H.~Sato, ``{NLP-based approaches for malware classification from API sequences},'' in \emph{IEEE Asia Pacific Symposium on Intelligent and Evolutionary Systems}, 2017.

\bibitem{vaswani2017attention}
A.~Vaswani, N.~Shazeer, N.~Parmar, J.~Uszkoreit, L.~Jones, A.~N. Gomez, {\L}.~Kaiser, and I.~Polosukhin, ``{Attention is all you need},'' \emph{Advances in neural information processing systems}, vol.~30, 2017.

\bibitem{lin2022survey}
T.~Lin and et~al., ``{A survey of transformers},'' \emph{AI Open}, 2022.

\bibitem{bracsoveanu2020visualizing}
A.~M. Bra{\c{s}}oveanu and R.~Andonie, ``{Visualizing transformers for nlp: a brief survey},'' in \emph{24th IEEE International Conference Information Visualisation}, 2020.

\bibitem{gillioz2020overview}
A.~Gillioz and et~al., ``{Overview of the Transformer-based Models for NLP Tasks},'' in \emph{15th IEEE Conference on Computer Science and Information Systems}, 2020.

\bibitem{park2022efficient}
H.~H. Park, Y.~Vyas, and K.~Shah, ``{Efficient classification of long documents using transformers},'' \emph{arXiv preprint}, 2022.

\bibitem{kenton2019bert}
J.~D. M.-W.~C. Kenton and et~al., ``{Bert: Pre-training of deep bidirectional transformers for language understanding},'' in \emph{Proceedings of naacL-HLT}, 2019.

\bibitem{achiam2023gpt}
J.~Achiam and et~al., ``{Gpt-4 technical report},'' \url{https://cdn.openai.com/papers/gpt-4.pdf}, 2023.

\bibitem{dosovitskiy2020image}
A.~Dosovitskiy and et~al., ``{An image is worth 16x16 words: Transformers for image recognition at scale},'' \emph{arXiv preprint}, 2020.

\bibitem{khan2022transformers}
S.~Khan and et~al., ``{Transformers in vision: A survey},'' \emph{ACM computing surveys}, 2022.

\bibitem{han2022survey}
K.~Han and et~al., ``{A survey on vision transformer},'' \emph{IEEE transactions on pattern analysis and machine intelligence}, 2022.

\bibitem{carion2020end}
N.~Carion and et~al., ``{End-to-end object detection with transformers},'' in \emph{European conference on computer vision, Springer}, 2020.

\bibitem{ramesh2022hierarchical}
A.~Ramesh and et~al., ``{Hierarchical text-conditional image generation with clip latents},'' \emph{arXiv preprint}, 2022.

\bibitem{videoworldsimulators2024}
T.~Brooks and et~al., ``{Video generation models as world simulators},'' \url{https://openai.com/research/video-generation-models-as-world-simulators}, 2024.

\bibitem{dong2018speech}
L.~Dong and et~al., ``{Speech-transformer: a no-recurrence sequence-to-sequence model for speech recognition},'' in \emph{IEEE international conference on acoustics, speech, signal processing}, 2018.

\bibitem{li2021mad}
M.~Q. Li and et~al., ``{I-MAD: Interpretable malware detector using galaxy transformer},'' \emph{Computers \& Security}, 2021.

\bibitem{fan2021heterogeneous}
Y.~Fan and et~al., ``{Heterogeneous temporal graph transformer: An intelligent system for evolving android malware detection},'' in \emph{Proceedings of the 27th ACM SIGKDD Conference on Knowledge Discovery \& Data Mining}, 2021.

\bibitem{dehunting}
A.~S. de~Oliveira and et~al., ``{Hunting Android Malware Using Multimodal Deep Learning and Hybrid Analysis Data},'' \emph{Brazilian Society of Computational Intelligence}, 2021.

\bibitem{lu2022research}
F.~Lu and et~al., ``{Research on the Construction of Malware Variant Datasets and Their Detection Method},'' \emph{Applied Sciences}, 2022.

\bibitem{li2023iot}
Y.~Li and Y.~Li, ``{IoT Malware Threat Hunting Method Based on Improved Transformer},'' \emph{International Journal of Network Security}, 2023.

\bibitem{ravi2023vit4mal}
A.~Ravi and et~al., ``{ViT4Mal: Lightweight Vision Transformer for Malware Detection on Edge Devices},'' \emph{ACM Transactions on Embedded Computing Systems}, 2023.

\bibitem{or2021pay}
O.~Or-Meir, A.~Cohen, Y.~Elovici, L.~Rokach, and N.~Nissim, ``{Pay attention: Improving classification of PE malware using attention mechanisms based on system call analysis},'' in \emph{IEEE International Joint Conference on Neural Networks}, 2021.

\bibitem{rahali2023malbertv2}
A.~Rahali and et~al., ``{MalBERTv2: Code Aware BERT-Based Model for Malware Identification},'' \emph{Big Data and Cognitive Computing}, 2023.

\bibitem{li2023gental}
L.~T. Li and et~al., ``{GenTAL: Generative Denoising Skip-gram Transformer for Unsupervised Binary Code Similarity Detection},'' in \emph{International Joint Conference on Neural Networks}, 2023.

\bibitem{hu2021single}
J.~L. Hu and et~al., ``{Single-shot black-box adversarial attacks against malware detectors: A causal language model approach},'' in \emph{IEEE International Conference on Intelligence and Security Informatics}, 2021.

\bibitem{ullah2022explainable}
F.~Ullah and et~al., ``{Explainable malware detection system using transformers-based transfer learning and multi-model visual representation},'' \emph{Sensors}, 2022.

\bibitem{zheng2013droid}
M.~Zheng and et~al., ``{Droid analytics: a signature based analytic system to collect, extract, analyze and associate android malware},'' in \emph{12th IEEE International Conference on Trust, Security and Privacy in Computing and Communications}, 2013.

\bibitem{chen2004model}
H.~Chen and et~al., ``{Model Checking One Million Lines of C Code.}'' in \emph{NDSS}.\hskip 1em plus 0.5em minus 0.4em\relax Citeseer, 2004.

\bibitem{damodaran2017comparison}
A.~Damodaran and et~al., ``{A comparison of static, dynamic, and hybrid analysis for malware detection},'' \emph{Journal of Computer Virology and Hacking Techniques}, 2017.

\bibitem{djenna2023artificial}
A.~Djenna and et~al., ``{Artificial Intelligence-Based Malware Detection, Analysis, and Mitigation},'' \emph{Symmetry}, 2023.

\bibitem{caviglione2020tight}
L.~Caviglione and et~al., ``{Tight arms race: Overview of current malware threats and trends in their detection},'' \emph{IEEE Access}, 2020.

\bibitem{kimmel2021recurrent}
J.~C. Kimmel and et~al., ``{Recurrent neural networks based online behavioural malware detection techniques for cloud infrastructure},'' \emph{IEEE Access}, 2021.

\bibitem{kimmell2021analyzing}
J.~C. Kimmell and et~al., ``{Analyzing machine learning approaches for online malware detection in cloud},'' in \emph{IEEE International Conference on Smart Computing (SMARTCOMP)}.\hskip 1em plus 0.5em minus 0.4em\relax IEEE, 2021.

\bibitem{gibert2020rise}
D.~Gibert and et~al., ``{The rise of machine learning for detection and classification of malware: Research developments, trends and challenges},'' \emph{Journal of Network \& Computer Applications}, 2020.

\bibitem{cortes1995support}
C.~Cortes and V.~Vapnik, ``{Support Vector Networks},'' \emph{Machine learning}, vol.~20, pp. 273--297, 1995.

\bibitem{rumelhart1986learning}
D.~E. Rumelhart, G.~E. Hinton, and R.~J. Williams, ``{Learning representations by back-propagating errors},'' \emph{nature}, 1986.

\bibitem{elman1990finding}
J.~L. Elman, ``{Finding structure in time},'' \emph{Cognitive science}, vol.~14, no.~2, pp. 179--211, 1990.

\bibitem{hochreiter1998vanishing}
S.~Hochreiter, ``{The vanishing gradient problem during learning recurrent neural nets and problem solutions},'' \emph{International Journal of Uncertainty, Fuzziness and Knowledge-Based Systems}, 1998.

\bibitem{hochreiter1997long}
S.~Hochreiter and J.~Schmidhuber, ``{Long short-term memory},'' \emph{Neural computation}, vol.~9, no.~8, pp. 1735--1780, 1997.

\bibitem{sutskever2014sequence}
I.~Sutskever, O.~Vinyals, and Q.~V. Le, ``{Sequence to sequence learning with neural networks},'' \emph{Advances in neural information processing systems}, vol.~27, 2014.

\bibitem{bahdanau2014neural}
D.~Bahdanau, K.~Cho, and Y.~Bengio, ``{Neural machine translation by jointly learning to align and translate},'' \emph{arXiv preprint}, 2014.

\bibitem{ba2016layer}
J.~L. Ba, J.~R. Kiros, and G.~E. Hinton, ``{Layer normalization},'' \emph{arXiv preprint arXiv:1607.06450}, 2016.

\bibitem{agarap2018deep}
A.~F. Agarap, ``{Deep learning using rectified linear units (relu)},'' \emph{arXiv preprint}, 2018.

\bibitem{hendrycks2016gaussian}
D.~Hendrycks and K.~Gimpel, ``{Gaussian error linear units (gelus)},'' \emph{arXiv preprint}, 2016.

\bibitem{sanh2019distilbert}
V.~a.~a. Sanh, ``{DistilBERT, a distilled version of BERT: smaller, faster, cheaper and lighter},'' \emph{arXiv preprint}, 2019.

\bibitem{liu2019roberta}
Y.~Liu and et~al., ``{Roberta: A robustly optimized bert pretraining approach},'' \emph{arXiv preprint}, 2019.

\bibitem{lan2019albert}
Z.~Lan and et~al., ``{Albert: A lite bert for self-supervised learning of language representations},'' \emph{arXiv preprint}, 2019.

\bibitem{clark2022canine}
J.~H. Clark and et~al., ``{Canine: Pre-training an efficient tokenization-free encoder for language representation},'' \emph{Transactions of the Association for Computational Linguistics}, 2022.

\bibitem{radford2019language}
A.~a.~a. Radford, ``{Language models are unsupervised multitask learners},'' \emph{OpenAI blog}, 2019.

\bibitem{chen2020android}
Y.-M. Chen and et~al., ``{Android malware detection system integrating block feature extraction and multi-head attention mechanism},'' in \emph{IEEE International Computer Symposium}, 2020.

\bibitem{hu2020exploit}
X.~Hu and et~al., ``{Exploit internal structural information for IoT malware detection based on hierarchical transformer model},'' in \emph{IEEE 19th International Conference on Trust, Security and Privacy in Computing and Communications}, 2020.

\bibitem{long2021detecting}
H.~Long and et~al., ``{Detecting Android Malware Based on Dynamic Feature Sequence and Attention Mechanism},'' in \emph{IEEE 5th International Conference on Cryptography, Security and Privacy}, 2021.

\bibitem{csahin2021malware}
N.~{\c{S}}ahin, ``{Malware detection using transformers-based model GPT-2},'' Master's thesis, Middle East Technical University, 2021.

\bibitem{wangwang2021network}
W.~Wangwang and et~al., ``{Network traffic oriented malware detection in IoT (internet-of-things)},'' in \emph{IEEE International Conference on Networking and Network Applications}, 2021.

\bibitem{bellante2021victory}
H.~Bellante and et~al., ``{Victory: A Framework For Fast Detection Of Iot Malware},'' \emph{Webology (ISSN: 1735-188X)}, 2021.

\bibitem{qi2022mdfa}
X.~Qi and et~al., ``{MDFA: A Malware Detection Framework Based on Attention Mechanism with Bi-LSTM},'' in \emph{4th IEEE International Conference on Applied Machine Learning}, 2022.

\bibitem{ghourabi2022security}
A.~Ghourabi, ``{A security model based on lightgbm and transformer to protect healthcare systems from cyberattacks},'' \emph{IEEE Access}, 2022.

\bibitem{seneviratne2022self}
S.~Seneviratne and et~al., ``{Self-supervised vision transformers for malware detection},'' \emph{IEEE Access}, 2022.

\bibitem{barut2022r1dit}
O.~Barut and et~al., ``{R1dit: Privacy-preserving malware traffic classification with attention-based neural networks},'' \emph{IEEE Transactions on Network and Service Management}, 2022.

\bibitem{pandya2023malware}
V.~Pandya and et~al., ``{Malware Detection through Contextualized Vector Embeddings},'' in \emph{IEEE Silicon Valley Cybersecurity Conference}, 2023.

\bibitem{deng2023transmalde}
X.~Deng and et~al., ``{TransMalDE: An Effective Transformer Based Hierarchical Framework for IoT Malware Detection},'' \emph{IEEE Transactions on Network Science and Engineering}, 2023.

\bibitem{saracino2023graph}
A.~Saracino and M.~Simoni, ``{Graph-Based Android Malware Detection and Categorization through BERT Transformer},'' in \emph{Proceedings of the 18th International Conference on Availability, Reliability and Security}, 2023.

\bibitem{jo2023malware}
J.~Jo and et~al., ``{A Malware Detection and Extraction Method for the Related Information Using the ViT Attention Mechanism on Android Operating System},'' \emph{Applied Sciences}, 2023.

\bibitem{piadatrans}
F.~Pi and et~al., ``{AdaTrans: An adaptive transformer for IoT Malware detection based on sensitive API call graph and inter-component communication analysis},'' \emph{Journal of Intelligent \& Fuzzy Systems}, 2023.

\bibitem{trizna2023nebula}
D.~Trizna and et~al., ``{Nebula: Self-Attention for Dynamic Malware Analysis},'' \emph{arXiv preprint}, 2023.

\bibitem{moon2021directional}
H.-J. Moon and et~al., ``{Directional Graph Transformer-Based Control Flow Embedding for Malware Classification},'' in \emph{Intelligent Data Engineering and Automated Learning: 22nd International Conference, IDEAL, Manchester, UK}.\hskip 1em plus 0.5em minus 0.4em\relax Springer, 2021.

\bibitem{rahali2021malbert}
A.~Rahali and et~al., ``{MalBERT: Malware Detection using Bidirectional Encoder Representations from Transformers},'' in \emph{IEEE International Conference on Systems, Man, and Cybernetics}, 2021.

\bibitem{li2022efficient}
C.~Li and et~al., ``{An Efficient Transformer Encoder-Based Classification of Malware Using API Calls},'' in \emph{IEEE 24th Int Conf on HPC \& Communications; 8th Int Conf on Data Science \& Systems; 20th Int Conf on Smart City; 8th Int Conf on Dependability in Sensor, Cloud \& Big Data Systems \& Application}, 2022.

\bibitem{chen2022malicious}
S.~Chen and et~al., ``{Malicious Code Family Classification Method Based on Vision Transformer},'' in \emph{IEEE 10th International Conference on Information, Communication and Networks}, 2022.

\bibitem{demirkiran2022ensemble}
F.~Demirk{\i}ran and et~al., ``{An ensemble of pre-trained transformer models for imbalanced multiclass malware classification},'' \emph{Computers \& Security}, 2022.

\bibitem{park2022vision}
K.-W. Park and S.-B. Cho, ``{A Vision Transformer Enhanced with Patch Encoding for Malware Classification},'' in \emph{International Conference on Intelligent Data Engineering and Automated Learning}.\hskip 1em plus 0.5em minus 0.4em\relax Springer, 2022.

\bibitem{belal2023global}
M.~M. Belal and D.~M. Sundaram, ``{Global-Local Attention-based Butterfly Vision Transformer for Visualization-based Malware Classification},'' \emph{IEEE Access}, 2023.

\bibitem{yang2022n}
C.~Yang and et~al., ``{N-Trans: Parallel Detection Algorithm for DGA Domain Names},'' \emph{Future Internet}, 2022.

\bibitem{gogoi2023dga}
B.~Gogoi and T.~Ahmed, ``{DGA domain detection using pretrained character based transformer models},'' in \emph{IEEE Guwahati Subsection Conference}, 2023.

\bibitem{guan2021malware}
Y.~Guan and et~al., ``{Malware system calls detection using hybrid system},'' in \emph{IEEE International Systems Conference}, 2021.

\bibitem{shahid2020devising}
S.~Shahid and et~al., ``{Devising Malware Characterstics using Transformers},'' \emph{arXiv preprint}, 2020.

\bibitem{demirci2022static}
D.~Dem{\i}rc{\i}, C.~Acarturk \emph{et~al.}, ``{Static malware detection using stacked BiLSTM and GPT-2},'' \emph{IEEE Access}, 2022.

\bibitem{hamad2021bertdeep}
S.~A. Hamad and et~al., ``{BERTDeep-Ware: A Cross-architecture Malware Detection Solution for IoT Systems},'' in \emph{IEEE 20th International Conference on Trust, Security and Privacy in Computing and Communications}, 2021.

\bibitem{bu2023triplet}
S.-J. Bu and S.-B. Cho, ``{Triplet-trained graph transformer with control flow graph for few-shot malware classification},'' \emph{Information Sciences}, 2023.

\bibitem{bu2020monte}
S.-J. Bu and et~al., ``{A Monte Carlo search-based triplet sampling method for learning disentangled representation of impulsive noise on steering gear},'' in \emph{IEEE International Conference on Acoustics, Speech and Signal Processing}, 2020.

\bibitem{guo2019star}
Q.~Guo and et~al., ``{Star-transformer},'' \emph{arXiv preprint}, 2019.

\bibitem{martin2019android}
A.~Mart{\'\i}n and et~al., ``{Android malware detection through hybrid features fusion and ensemble classifiers: The AndroPyTool framework and the OmniDroid dataset},'' \emph{Information Fusion}, 2019.

\bibitem{freitas2022malnet}
S.~Freitas and et~al., ``{MalNet: A large-scale image database of malicious software},'' in \emph{Proceedings of the 31st ACM International Conference on Information \& Knowledge Management}, 2022.

\bibitem{pa2015iotpot}
Y.~M.~P. Pa and et~al., ``{IoTPOT: Analysing the Rise of IoT Compromises},'' in \emph{9th USENIX Workshop on Offensive Technologies (WOOT 15)}, 2015.

\bibitem{pa2016iotpot}
Y.~M. Pa and et~al., ``{IoTPOT: A novel honeypot for revealing current IoT threats},'' \emph{Journal of Information Processing}, 2016.

\bibitem{garcia2020iot23}
S.~Garcia and et~al., ``{{IoT-23: A labeled dataset with malicious and benign IoT network traffic}},'' \url{http://doi.org/10.5281/zenodo.4743746}, 2020.

\bibitem{malgenomeprojectAndroidMalware}
``{{A}ndroid {M}alware {G}enome {P}roject --- malgenomeproject.org},'' \url{http://www.malgenomeproject.org/}, [Accessed on Feb15, 2024].

\bibitem{taheri2019extensible}
L.~Taheri and et~al., ``{Extensible android malware detection and family classification using network-flows and API-calls},'' in \emph{IEEE International Carnahan Conference on Security Technology}, 2019.

\bibitem{ahmed2021ecu}
M.~Ahmed and et~al., ``{ECU-IoHT: A dataset for analyzing cyberattacks in Internet of Health Things},'' \emph{Ad Hoc Networks}, 2021.

\bibitem{booij2021ton_iot}
T.~M. Booij and et~al., ``{ToN\_IoT: The role of heterogeneity and the need for standardization of features and attack types in IoT network intrusion data sets},'' \emph{IEEE Internet of Things Journal}, 2021.

\bibitem{ferrag2022edge}
M.~A. Ferrag and et~al., ``{Edge-IIoTset: A new comprehensive realistic cyber security dataset of IoT and IIoT applications for centralized and federated learning},'' \emph{IEEE Access}, 2022.

\bibitem{anderson2018ember}
H.~S. Anderson and P.~Roth, ``{Ember: an open dataset for training static pe malware machine learning models},'' \emph{arXiv preprint}, 2018.

\bibitem{fernandez2019intelligent}
L.~Fernandez~Maimo and et~al., ``{Intelligent and dynamic ransomware spread detection and mitigation in integrated clinical environments},'' \emph{Sensors}, 2019.

\bibitem{kaggle2021iotmalware}
Kaggle, ``{{IOT\_Malware\_dataset\_for Classification}},'' \url{https://www.kaggle.com/datasets/anaselmasry/iot-malware}, 2021, [Accessible on Feb 15, 2024].

\bibitem{arp2014drebin}
D.~Arp and et~al., ``{Drebin: Effective and explainable detection of android malware in your pocket.}'' in \emph{Ndss}, 2014.

\bibitem{allix2016androzoo}
K.~Allix and et~al., ``{Androzoo: Collecting millions of android apps for the research community},'' in \emph{Proceedings of the 13th international conference on mining software repositories}, 2016.

\bibitem{mahdavifar2020dynamic}
S.~Mahdavifar and et~al., ``{Dynamic android malware category classification using semi-supervised deep learning},'' in \emph{IEEE Intl Conf on Dependable, Autonomic and Secure Computing, Intl Conf on Pervasive Intelligence and Computing, Intl Conf on Cloud and Big Data Computing, Intl Conf on Cyber Science and Technology Congress}, 2020.

\bibitem{catak2019benchmark}
F.~O. Catak and A.~F. Yaz{\i}, ``{A benchmark API call dataset for windows PE malware classification},'' \emph{arXiv preprint}, 2019.

\bibitem{harang2020sorel}
R.~Harang and E.~M. Rudd, ``{SOREL-20M: A large scale benchmark dataset for malicious PE detection},'' \emph{arXiv preprint}, 2020.

\bibitem{de2023behavioral}
A.~S. de~Oliveira and et~al., ``{Behavioral malware detection using deep graph convolutional neural networks},'' \emph{Authorea Preprints}, 2023.

\bibitem{nataraj2011malware}
L.~Nataraj and et~al., ``{Malware images: visualization and automatic classification},'' in \emph{Proceedings of the 8th international symposium on visualization for cyber security}, 2011.

\bibitem{004e-v304-19}
A.~Oliveira, ``{Malware Analysis Datasets: Top-1000 PE Imports},'' \url{https://dx.doi.org/10.21227/004e-v304}, 2019.

\bibitem{speakeasy2021}
``{Speakeasy: portable, modular, binary emulator designed to emulate Windows kernel and user mode malware},'' \url{https://github.com/mandiant/speakeasy}, Nov. 2021, [Accessed on Feb15, 2024].

\bibitem{bosansky2022avast}
B.~Bosansky, D.~Kouba, O.~Manhal, T.~Sick, V.~Lisy, J.~Kroustek, and P.~Somol, ``{Avast-{CTU} public {CAPE} dataset},'' \emph{arXiv preprint}, 2022.

\bibitem{mcd2019}
``{Malicious Code Dataset},'' \url{https://github.com/kericwy1337/Datacon2019-Malicious-Code-DataSet-Stage1}, Jul. 2019, [Accessed on Feb15, 2024].

\bibitem{virustotal2021}
{Chronicle LLC}, ``{VirusTotal},'' \url{https://www.virustotal.com/gui/}[ Accessed on Feb15, 2024].

\bibitem{virusshare}
``{VirusShare},'' \url{https://virusshare.com/}, [Accessed Feb15, 2024].

\bibitem{ghodke2018alexa}
S.~Ghodke, ``{{Alexa Top 1 million sites}},'' \url{https://www.kaggle.com/datasets/cheedcheed/top1m}, [Accessed on Feb15, 2024].

\bibitem{ExpiredDomainsAlexaTop}
``{Alexa Top Websites},'' \url{https://www.expireddomains.net/alexa-top-websites/}, [Accessed on Feb15, 2024].

\bibitem{dganetlab2022}
``{DGA Dataset},'' \url{https://data.netlab.360.com/dga/}, [Accessed by respective author on 10 December 2021].

\bibitem{stratosphere2015datasets}
``{Stratosphere Laboratory Datasets},'' \url{https://www.stratosphereips.org/datasets-overview}, 2015, [Accessed on Feb15, 2024].

\bibitem{sharafaldin2018toward}
I.~Sharafaldin and et~al., ``{Toward generating a new intrusion detection dataset and intrusion traffic characterization.}'' \emph{ICISSp}, 2018.

\bibitem{netlabopendata2022}
``{{Home - Netlab OpenData Project}},'' \url{https://data.netlab.360.com/feeds/dga/dga.txt}, accessed by author on Jun29, 2022.

\bibitem{abuseMalwareBazaarMalware}
``{{M}alware{B}azaar | {M}alware sample exchange},'' \url{https://bazaar.abuse.ch/}, [Accessed on Feb15, 2024].

\bibitem{fraunhoferMalpediaFraunhofer}
F.~FKIE, ``{{M}alpedia},'' \url{https://malpedia.caad.fkie.fraunhofer.de/}, [Accessed on Feb15, 2024].

\bibitem{phandi2018semeval}
P.~Phandi and et~al., ``{SemEval-2018 task 8: Semantic extraction from CybersecUrity REports using natural language processing (SecureNLP)},'' in \emph{Proceedings of The 12th International Workshop on Semantic Evaluation}, 2018.

\bibitem{ronen2018microsoft}
R.~Ronen, M.~Radu, C.~Feuerstein, E.~Yom-Tov, and M.~Ahmadi, ``{Microsoft malware classification challenge},'' \emph{arXiv preprint}, 2018.

\bibitem{haddadpajouh2018deep}
H.~HaddadPajouh and et~al., ``{A deep recurrent neural network based approach for internet of things malware threat hunting},'' \emph{Future Generation Computer Systems}, 2018.

\bibitem{dymshits2017process}
M.~Dymshits and et~al., ``{Process monitoring on sequences of system call count vectors},'' in \emph{IEEE International Carnahan Conference on Security Technology}, 2017.

\bibitem{googleAndroidApps}
``{{A}ndroid {A}pps on {G}oogle {P}lay},'' \url{https://play.google.com/store}, [Accessed on Feb15, 2024].

\bibitem{Kim2018PEHeader}
S.~Kim, ``{{PE} {H}eader {A}nalysis for {M}alware {D}etection},'' \url{https://scholarworks.sjsu.edu/etd_projects/624}, 2018.

\bibitem{perozzi2014deepwalk}
B.~Perozzi and et~al., ``Deepwalk: Online learning of social representations,'' in \emph{Proceedings of the 20th ACM SIGKDD international conference on Knowledge discovery and data mining}, 2014.

\bibitem{mikolov2013efficient}
T.~Mikolov and et~al., ``{Efficient estimation of word representations in vector space},'' \emph{arXiv preprint}, 2013.

\bibitem{nagrani2021attention}
A.~Nagrani and Y.~et~al., ``Attention bottlenecks for multimodal fusion,'' \emph{Advances in neural information processing systems}, 2021.

\bibitem{zhang2023cross}
J.~Zhang, Y.~Xie, W.~Ding, and Z.~Wang, ``Cross on cross attention: Deep fusion transformer for image captioning,'' \emph{IEEE Transactions on Circuits and Systems for Video Technology}, 2023.

\bibitem{chen2021decision}
L.~Chen and et~al., ``Decision transformer: Reinforcement learning via sequence modeling,'' \emph{Advances in neural information processing systems}, vol.~34, pp. 15\,084--15\,097, 2021.

\bibitem{chen2022transdreamer}
C.~Chen, Y.-F. Wu, J.~Yoon, and S.~Ahn, ``Transdreamer: Reinforcement learning with transformer world models,'' \emph{arXiv preprint arXiv:2202.09481}, 2022.

\bibitem{garg2020bae}
S.~Garg and G.~Ramakrishnan, ``Bae: Bert-based adversarial examples for text classification,'' \emph{arXiv preprint}, 2020.

\bibitem{guo2021gradient}
C.~Guo and et~al., ``Gradient-based adversarial attacks against text transformers,'' \emph{arXiv preprint arXiv:2104.13733}, 2021.

\bibitem{liu2022character}
A.~Liu and et~al., ``Character-level white-box adversarial attacks against transformers via attachable subwords substitution,'' \emph{arXiv preprint}, 2022.

\bibitem{wei2022towards}
Z.~Wei and et~al., ``Towards transferable adversarial attacks on vision transformers,'' in \emph{Proceedings of the AAAI Conference on Artificial Intelligence}, 2022.

\bibitem{gao2024pe}
S.~Gao and et~al., ``Pe-attack: On the universal positional embedding vulnerability in transformer-based models,'' \emph{IEEE Transactions on Information Forensics and Security}, 2024.

\end{thebibliography}

\vspace{-25mm}
\begin{IEEEbiography}[{\includegraphics[width=1.1in,height=1.3in,clip,keepaspectratio]{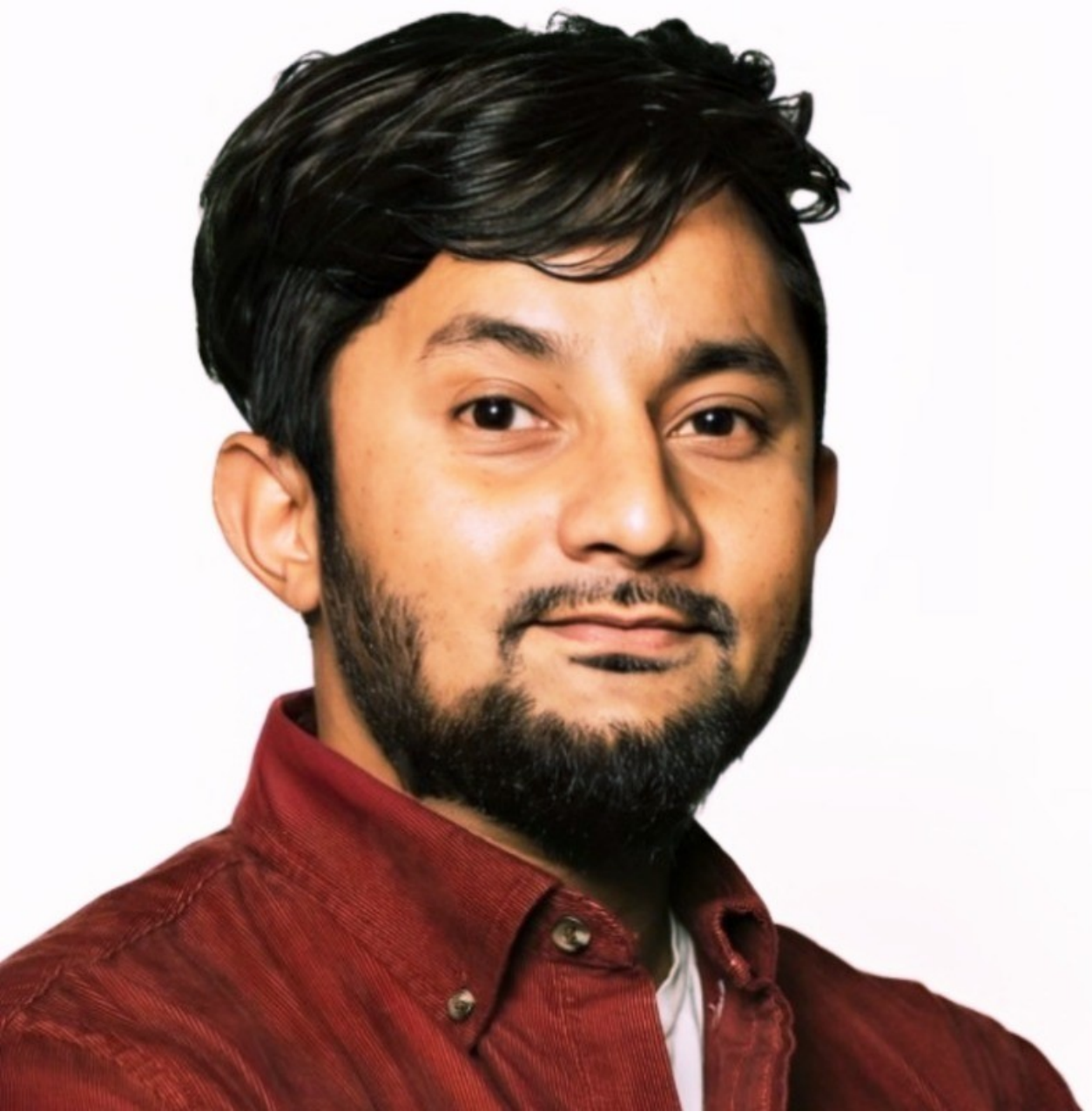}}]{Pradip Kunwar} is currently pursuing Ph.D. in Computer Science at Tennessee Tech University, Cookeville, TN, USA. He received his B.Tech degree in Electronics and Communication from NIT Rourkela, India. His current research interests are Transformers in Malware Analysis, Security of/for LLMs, PEFT Methods, Mixture of Experts Architectures etc. He is interested in researching the underlying vulnerabilities of AI systems and making them more robust against adversarial attacks. 
\end{IEEEbiography}
\vspace{-11 mm}

\begin{IEEEbiography}[{\includegraphics[width=1.1in,height=1.1
in]{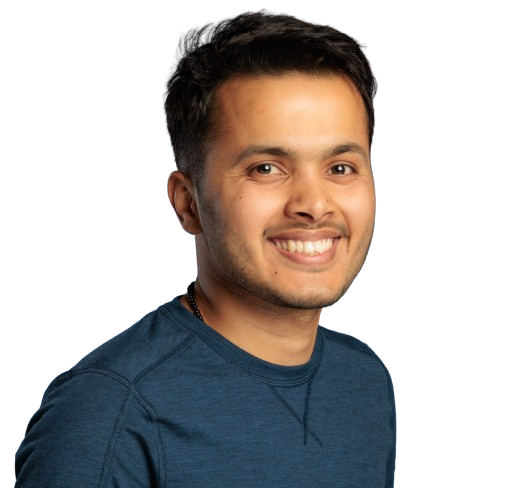}}]{Kshitiz Aryal}
 is currently pursuing a Ph.D. degree with the Department of Computer Science, Tennessee Technological University, Cookeville, TN, USA. He also received his M.S. in Computer Science from Tennessee Tech and B.S. in ECE from Tribhuvan University, Nepal.  His current research interests include adversarial attacks/defense, malware analysis, AI security, security of AI, explainable AI, and data science.  
\end{IEEEbiography}

\vspace{-11 mm}

\begin{IEEEbiography}[{\includegraphics[width=1.1in,height=1.3in]{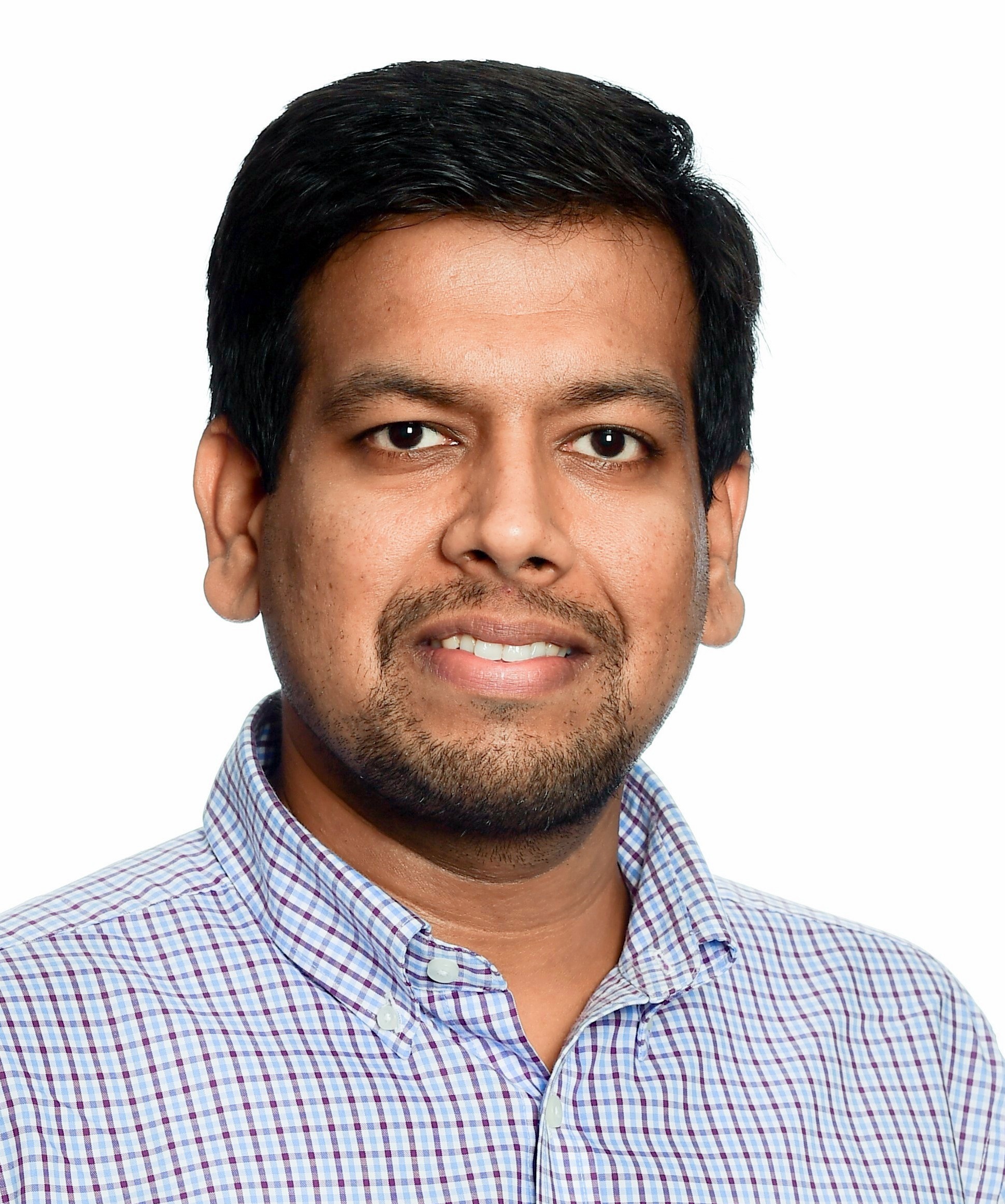}}]{Maanak Gupta} (Senior Member, IEEE) is an Assistant Professor in Computer Science at Tennessee Technological University, Cookeville, USA. He received M.S. and Ph.D. in Computer Science from the University of Texas at San Antonio (UTSA) and has also worked as a postdoctoral fellow at the Institute for Cyber Security (ICS) at UTSA. His primary area of research includes security and privacy in cyber space focused in studying foundational aspects of access control, malware analysis, AI and machine learning assisted cyber security, and their applications in technologies including cyber physical systems, cloud computing, IoT and Big Data. His research has been funded by the US National Science Foundation (NSF), NASA, and US Department of Defense (DoD) among others.
\end{IEEEbiography}

\vspace{-11 mm}

\begin{IEEEbiography}[{\includegraphics[width=1.1in,height=1.3in,clip,keepaspectratio]{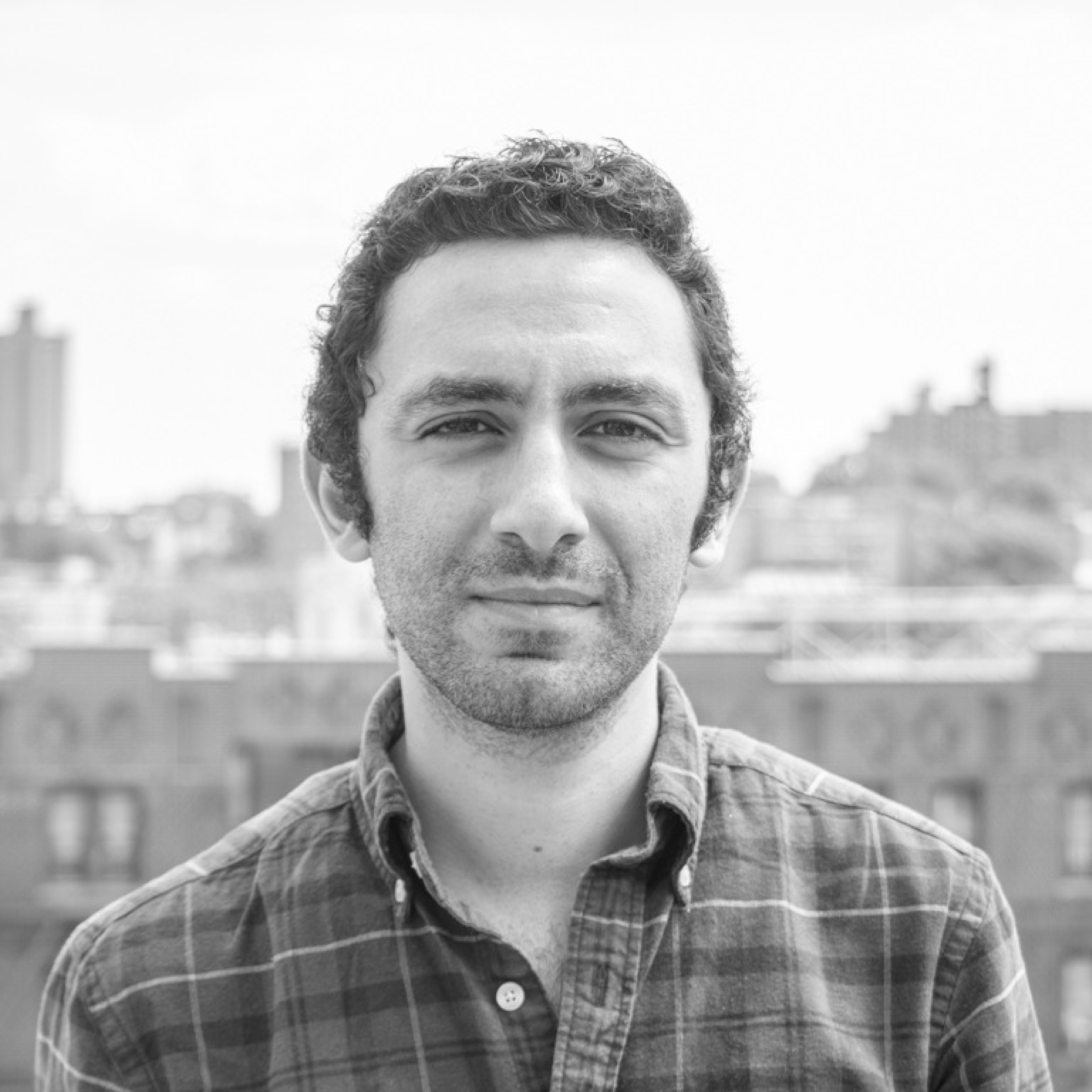}}]{Mahmoud Abdelsalam}
received the M.Sc. and Ph.D. degrees from the University of Texas at San Antonio (UTSA), in 2017 and 2018, respectively. He was working as a Postdoctoral Research Fellow with the Institute for Cyber Security (ICS), UTSA, and as an Assistant Professor with the Department of Computer Science, Manhattan College. He is currently working as an Assistant Professor with the Department of Computer Science, North Carolina A\&T State University. His research interests include computer systems security, anomaly and malware detection, cloud computing security and monitoring, cyber-physical systems security, and applied ML.
\end{IEEEbiography}
\vspace{-10 mm}

\begin{IEEEbiography}[{\includegraphics[width=1.1in,height=1.3in,clip,keepaspectratio]{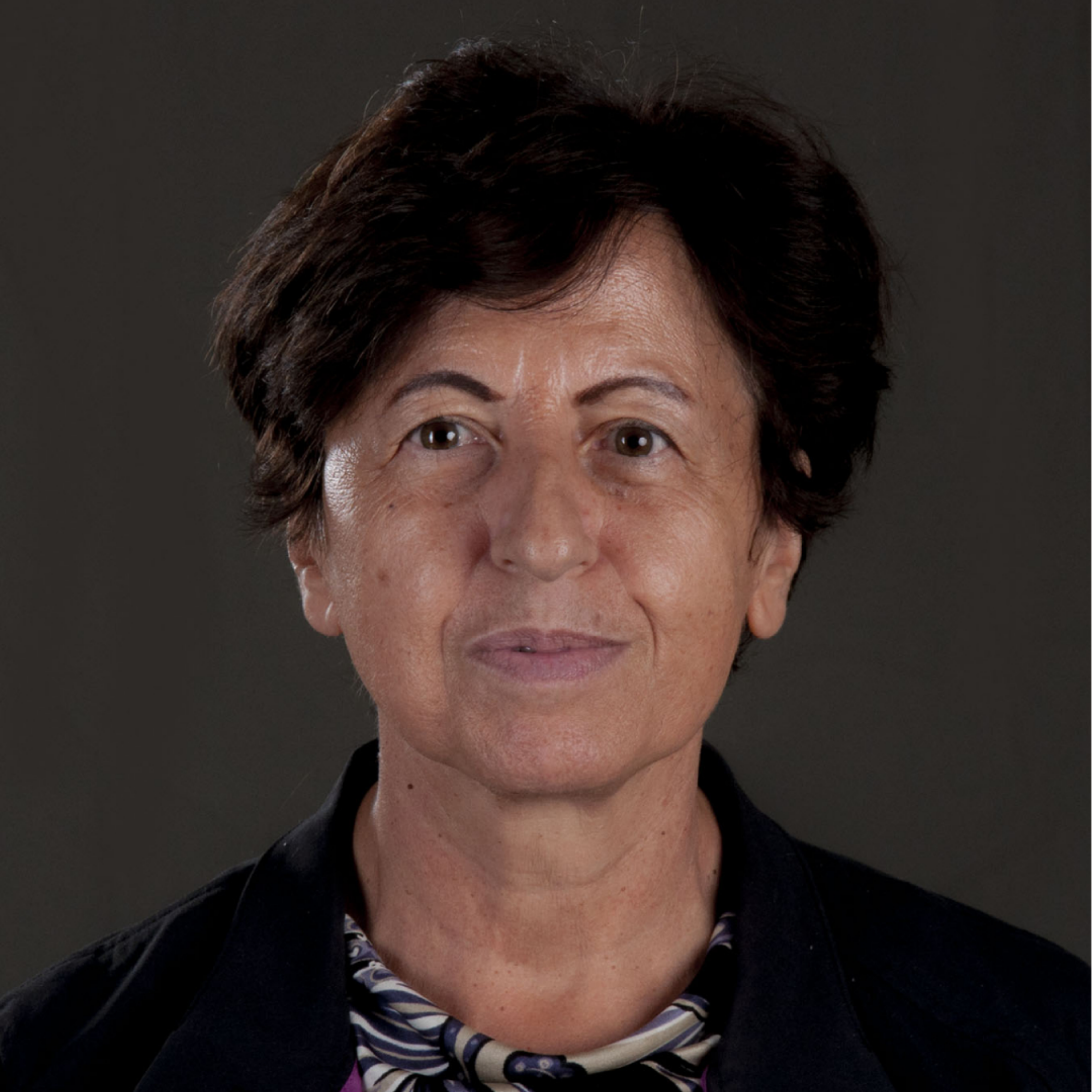}}]{Elisa Bertino} is Samuel Conte professor of Computer Science at Purdue University. Prior to joining Purdue, she was a professor and department head at the Department of Computer Science and Communication of the University of Milan. She has been a visiting researcher at the IBM Research Laboratory in San Jose (now Almaden), at Rutgers University, at Telcordia Technologies. She also held visiting professor positions at the Singapore National University and the Singapore Management University. Her recent research focuses on security and privacy of cellular networks and IoT systems, and on edge analytics for cybersecurity.  Elisa Bertino is a Fellow member of IEEE, ACM, and AAAS. She received the 2002 IEEE Computer Society Technical Achievement Award for “For outstanding contributions to database systems and database security and advanced data management systems”, the 2005 IEEE Computer Society Tsutomu Kanai Award for “Pioneering and innovative research contributions to secure distributed systems”, the 2019-2020 ACM Athena Lecturer Award, and the 2021 IEEE 2021 Innovation in Societal Infrastructure Award. She received an Honorary Doctorate from Aalborg University in 2021 and an Honorary Research Doctorate in Computer Science from the University of Salerno in 2023. She is currently serving as ACM Vice-president.

\end{IEEEbiography}


\vfill


\section{Appendix}
\subsection{Technical Comparison - Transformers Adaptation and  Effectiveness in Malware Analysis}
\label{technicaladaption}
In this section, we discuss
modifications for adapting the models presented to the malware analysis domain and the effectiveness of these modifications. In our review of transformer-based approaches to malware analysis, we observed that specific transformer architectures like BERT, GPT-2, and ViT are frequently employed due to their ability to handle different types of data representations, such as sequences, text, and images. Furthermore, we explore how researchers have customized these models, such as the introduction of additive attention mechanisms and residual weight parameters to reduce computational complexity in models tailored for resource-constrained environments. These modifications not only enhance the efficiency of the models but also allow them to better capture the intricate patterns of malware behavior.

\begin{table*}[hb!]
\centering
\caption{Malware Analysis: Comprehension of Modeling Techniques, Unique Contributions, and Effectiveness}
\label{Table:OthersTable}
\vspace{-3mm}
\begin{adjustbox}{width=\textwidth}
\begin{tabular}{|p{3cm}|p{4cm}|p{7cm}|p{9cm}|}
\hline
\rowcolor{gray!15}
\textbf{Transformer} & \textbf{Task} & \textbf{Modifications / Unique Contribution}  & \textbf{Effectiveness}\\
\hline
BERT, XLNet, and ULMFiT [2020] ~\cite{shahid2020devising}&	Malware characteristic classification&\textbf{Fine-tuned} on a cybersecurity dataset&BERT and XLNET showed promise in \textbf{handling context and complexity} of malware-related sentences\\
\hline
\rowcolor{gray!10}
Hybrid LSTM Transformer [2021]~\cite{guan2021malware}&	Anomaly Detection&\textbf{Combines} LSTM and Transformer for enhanced feature extraction& Demonstrates the effectiveness for \textbf{temporal dynamics and global attention} with 1.6\% improvement compared to LSTM-based models\\
\hline
GPT-2 Transformer [2021]~\cite{hu2021single}	&Single-shot black-box adversarial evasion&	\textbf{Fine-tuned} GPT-2 on benign file sequences to generate benign-looking adversarial perturbations&	Created adversarial perturbations in a single query, achieving \textbf{24.51\% evasion rate} without multiple feedback loops\\
\hline
\rowcolor{gray!10}
BERT Transformer [2022]~\cite{ghourabi2022security}&	Intrusion and Malware detection in healthcare systems&\textbf{Fine-tuned} on textual network flow data from healthcare, LightGBM for malware and intrusion detection&Hybrid approach leverages transformer and gradient-boosting to \textbf{optimize performance and memory efficiency}\\
\hline
Parallel N-gram Transformer [2022] ~\cite{yang2022n}&	Detection of DGA domains&Parallel combination of the N-gram algorithm and \textbf{Transformer to enhance feature extraction}&	Parallel N-gram architecture ensures richer feature extraction achieving an \textbf{accuracy of 96.97}\%\\
\hline
\rowcolor{gray!10}
Residual 1D Image Transformer [2022]~\cite{barut2022r1dit}&	Privacy-preserving malware traffic classification& Uses \textbf{1-D convolutions \& 2-D positional encoding.} Meta-learning done using transfer \& few-shot learning&	Dimensional changes \textbf{improved the feature extractions} improving the detection rate of new and unseen malware\\
\hline

CANINE-c [2023]~\cite{gogoi2023dga}	&Detection of DGA domains& \textbf{Fine-tuned} on 1 million DGA and benign domain names&Fine-tuning the character-level processing enhances the models \textbf{accuracy to 99}\%\\
\hline
\rowcolor{gray!10}
GenTAL, skip-gram with denoising [2023]~\cite{li2023gental}&	Unsupervised binary code similarity detection& \textbf{Pre-trained} using masked instruction recovery, \textbf{Fine-tuned}, Combines skip-gram loss, denoising autoencoder principles using the transformer &Better handles semantic variations in binary code due to different compilation options, achieved\textbf{ 0.76 and 0.74 MRR} for cross-compiler \& obfuscation detection\\
\hline

\end{tabular}
\end{adjustbox}
\end{table*}

\begin{table*}[hb!]
\centering
\caption{IoT Malware Analysis: Comprehension of Modeling Techniques, Unique Contributions, and Effectiveness}
\label{Table:IoTTable}
\vspace{-3mm}
\begin{adjustbox}{width=\textwidth}
\begin{tabular}{|p{3.5cm}|p{4cm}|p{7cm}|p{9cm}|}
\hline
\rowcolor{gray!15}
\textbf{Transformer} & \textbf{Task} & \textbf{Modifications / Unique Contribution}  & \textbf{Effectiveness}\\
\hline

Hierarchical Transformer [2020]\cite{hu2020exploit}&	IoT Malware Detection&Leverages internal information from malware with two-layer hierarchy \& \textbf{integrates multi-head attention}& Hierarchical design helps to better understand packet \& session based anomalies, \textbf{significantly improves} detection \textbf{accuracy to 98.75}\%\\
\hline
\rowcolor{gray!10}
Vanilla Transformer [2021]\cite{wangwang2021network}&	IoT malware detection&Transformer was applied to analyze network traffic features and \textbf{enhance representations}&\textbf{Improved detection accuracy} to 99.38\% accuracy on character-level features and 99.52\% accuracy on word-level features\\
\hline
BERT based VICTORY [2021]\cite{bellante2021victory}&	IoT malware detection&BERT fine-tuned using embeddings from API calls focusing on \textbf{rapid re-training with minimal data}&	Fine-tuning with BERT helped maintain \textbf{high accuracy even with low malware sample rates}\\
\hline
\rowcolor{gray!10}
BERT + CNN-BiLSTM + Local Attention [2021]\cite{hamad2021bertdeep}&	Cross-architecture IoT malware detection& BERTDeep-Ware enhances feature representations, \textbf{Fine-tuned} for multiple IoT CPU architectures&Significantly \textbf{improved detection} of \textbf{complex malware behaviors} with more than 90\% accuracy\\
\hline

Custom model (TransMalDE) [2023]\cite{deng2023transmalde}&	IoT malware detection & Integrates Transformer, uses \textbf{location embedding} to model sensitive API subgraphs&Demonstrated \textbf{superior performance} compared to CNN, LSTM, and GRU, with over 96\% accuracy\\
\hline
\rowcolor{gray!10}
Transformer with additive attention, residual weight parameters [2023]\cite{li2023iot}&	IoT malware threat hunting&	Improved by \textbf{replacing multi-head with additive attention}, added \textbf{residual weight parameters} to tackle vanishing gradients&Additive attention \textbf{scales linearly with the sequence length without compromising accuracy}\\
\hline
ViT4Mal lightweight ViT [2023]\cite{ravi2023vit4mal}&	Malware detection on edge devices&\textbf{Optimized for low resource usage}, with quantization and hardware-specific optimizations&ViT4Mal achieves \textbf{competitive accuracy with minimal layers} and attention heads and \textbf{41x speedup} compared to the original ViT\\
\hline
\rowcolor{gray!10}
Adaptive Transformer with an adaptive factor (AdaTrans) [2023]\cite{piadatrans}&IoT malware detection&	AdaTrans \textbf{enhances} the standard Transformer allowing to\textbf{ weigh the importance of different semantic features}&	AdaTrans’ adaptive multi-head attention mechanism proved \textbf{highly effective in capturing subtle malware behaviors} that exploit Android’s ICC (Inter Component Communication) system\\
\hline

\end{tabular}
\end{adjustbox}
\end{table*}


In Tables~\ref{Table:OthersTable}, \ref{Table:IoTTable}, \ref{Table:WindowsTable}, and \ref{Table:AndroidTable}  we list the type of transformer integrated to the work, the task carried out as the goal of the work, the modifications performed, the unique adaptation methods implemented in the approach, and the effectiveness of the modifications. Since the details of the above-mentioned aspects for the papers describing those approaches are content-heavy, we make our analysis as concise as possible to enable the readers to understand the work with minimal context without losing the underlying knowledge.  

Table~\ref{Table:OthersTable} summarizes our analysis of  approaches designed for tasks, other than malware detection and classification, such as detection evasion, malicious domain name detection, binary code similarity detection, anomaly detection, privacy-preserving malware classification, and malware characteristics classification. Table~\ref{Table:IoTTable},  summarizes our analysis of  approaches designed for the detection and classification of \textbf{IoT malware system}. Since the resource and infrastructure constraints in IoT systems are typical limited, we found several approaches focusing on addressing the computational limitations in the IoT system while integrating the abilities of transformers in their system.

Table~\ref{Table:WindowsTable} summarizes our analysis of  approaches designed the detection and classification of malware performed on the \textbf{Windows malware system}. Our analysis shows that there have been numerous approaches focusing on the design of robust malware detectors using diverse sets of features and modifications in the existing workflow of transformers as well as implementations of various methods like fine tuning, multi-modal integration, etc. 
Table \ref{Table:AndroidTable}, summarizes our analysis of  approaches designed for the detection and classification of \textbf{Android malware system}. We observed that many approaches integrate transformers to harness sequential features and images. Besides, an interesting approach, using multiple transformer models to map the evolution and propagation patterns of malware, also shows promising results in the use of transformers. 

\begin{table*}[ht!]
\centering
\caption{Windows Malware Analysis: Comprehension of Modeling Techniques, Unique Contributions, and Effectiveness}
\label{Table:WindowsTable}
\vspace{-3mm}
\begin{adjustbox}{width=\textwidth}
\begin{tabular}{|p{3.5cm}|p{4cm}|p{7cm}|p{9cm}|}
\hline
\rowcolor{gray!15}
\textbf{Transformer} & \textbf{Task} & \textbf{Modifications / Unique Contribution}  & \textbf{Effectiveness}\\
\hline
Directional Graph Transformer (DGT) [2021]~\cite{moon2021directional}&Malware classification&DeepWalk algorithm converts Control Flow Graphs into sequences for Transformer to \textbf{generate embeddings}&Framework \textbf{results significant improvements in detecting malware that uses complex control flow}\\
\hline
\rowcolor{gray!10}
GPT-2 [2021]~\cite{csahin2021malware}&	Malware Detection&	\textbf{Fine-tuned} GPT-2 on assembly sequences from PE files&Fine-tuning resulted\textbf{ better accuracy} than not fine tuning\\
\hline

LSTM Attention \& Transformer [2021]~\cite{or2021pay}&	Malware Classification&	Experimented\textbf{ Transformer integrating LSTM} with attentions and without&\textbf{For shorter sequences, LSTM} with attention outperformed Transformers in both accuracy, training time and \textbf{opposite for longer sequences}\\
\hline
\rowcolor{gray!10}
\textbf{I-MAD:} Galaxy Transformer [2021]~\cite{li2021mad}&	Malware Detection&	Introduced \textbf{3 layer hierarchical Galaxy Transformer} with Satellite-Planet, Planet-Star, and Star-Galaxy transformers to model malware's structure at multiple levels&	\textbf{Reduced time-space complexity} to O(n), enabled model to \textbf{process very long sequences} - previously unmanageable with standard transformer, \textbf{better robustness} for generalization \textbf{to unseen malware}\\
\hline

Vision Transformer (ViT) with Lambda Attention [2022] ~\cite{chen2022malicious}&	Malware Classification&\textbf{Replaced self-attention quadratic computations with linear Lambda Layer} by introducing positional relationship learning using $\lambda_c$ (contextual) and $\lambda_{np}$ (positional)&Lambda Layer\textbf{ reduces memory complexity} while maintaining lower resource consumption and \textbf{reducing training time by up to 30x}\\
\hline
\rowcolor{gray!10}
GPT-2 with Stacked BiLSTM [2022]~\cite{demirci2022static}&	Malware Detection&GPT-2 \textbf{fine-tuned} on assembly instructions with\textbf{ stacked BiLSTM layers} for document \& sentence-level analysis&Combination \textbf{enhances contextual understanding} and improves detection performance\\
\hline

ViT enhanced with patch encoding [2022]~\cite{park2022vision}	&Malware Classification	&\textbf{Fine-tuned} ViT employing \textbf{data augmentation with patch encoded malware images}&	Fine-tuning helps in \textbf{capturing patterns missed by traditional models}\\
\hline
\rowcolor{gray!10}
Transformer Encoder [2022]~\cite{li2022efficient}	&Malware Classification&MalTransEn leverages \textbf{self-attention} mechanisms for classifying API call sequences&MalTransEn's multi-head self-attention \textbf{efficiently captures interactions} in API call sequences\\
\hline

BERT-based model [2022]~\cite{lu2022research}	&Malware Variant Construction and Detection&	\textbf{Pre-trained} on malware API sequences \& \textbf{fine-tuned } by adversarial training for robustness against obfuscation&Adversarial training resulted in \textbf{better} model \textbf{at handling obfuscated API sequences}\\
\hline
\rowcolor{gray!10}
\textbf{MDFA}: Bi-LSTM with Attention [2022] ~\cite{qi2022mdfa}&	Malware Detection&\textbf{Capture temporal dependencies} in API sequences&	Attention allowed model to \textbf{focus on critical API calls} within long sequences outperforming traditional models\\
\hline

BERT-CANINE Ensemble [2022]~\cite{demirkiran2022ensemble}&Imbalanced Multiclass Malware Classification&\textbf{Introduced} Random Transformer Forest, \textbf{bagging-based ensemble}, with \textbf{fine-tuning} on API call sequences&Ensemble \textbf{improves classification} on \textbf{highly imbalanced datasets}\\
\hline
\rowcolor{gray!10}
BERT, DistilBERT, AlBERT, RoBERTa [2023]~\cite{pandya2023malware}&	Malware Classification	&Generate \textbf{context-aware} opcode embeddings&Combining contextual embeddings with Resnet-18 CNN results \textbf{applicability of transformers in conjunction with deep learning} classifiers\\
\hline

Graph Transformer with triplet loss [2023]~\cite{bu2023triplet}&	Few-shot Malware Classification	&\textbf{Introduction of graph transformer} to embed malware control flow graphs (CFGs) generated through \textbf{triplet loss} used with Graph transformer &Integration of triplet loss resulted in significant improvements in classification with limited samples and \textbf{demonstrates model's effectiveness in few-shot settings}\\
\hline
\rowcolor{gray!10}
\textbf{B\_ViT} [2023]~\cite{belal2023global}&	Malware Classification&	\textbf{1.} B\_ViT (with Global-Local Attention, Adaptive Positional Encoding) integrates four stages: image partitioning, local attention, global attention, \& classification. Also, enables parallel processing& Significantly \textbf{reduced training time}, with a speed-up factor of 2.42 over IEViT and 1.81 over ViT, \textbf{improved detection} accuracy for \textbf{obfuscated and polymorphic} malware\\
\hline

\textbf{Nebula} [2023]~\cite{trizna2023nebula}	&Malware Detection and Classification&Nebula \textbf{utilizes a self-attention-based} transformer model to process dynamic behavioral reports&Self-attention mechanism allowed Nebula to \textbf{generalize better} across dynamic logs, improving detection accuracy\\
\hline

\end{tabular}
\end{adjustbox}
\end{table*}

\begin{table*}[ht!]
\centering
\caption{Android Malware Analysis: Comprehension of Modeling Techniques, Unique Contributions, and Effectiveness}
\vspace{-3mm}
\label{Table:AndroidTable}
\begin{adjustbox}{width=\textwidth}
\begin{tabular}{|p{3.5cm}|p{4cm}|p{7cm}|p{9cm}|}
\hline
\rowcolor{gray!15}
\textbf{Transformer} & \textbf{Task} & \textbf{Modifications / Unique Contribution}  & \textbf{Effectiveness}\\

\hline
Transformer and Bi-LSTM [2020]~\cite{chen2020android}&Android malware detection&\textbf{Multi-model architecture} using block feature extraction combined with Multi-Head Attention&Combination of block feature extraction and Multi-Head Attention allowed the model to \textbf{efficiently classify malware}\\
\hline

\rowcolor{gray!10}
Transformer model with attention [2021]~\cite{long2021detecting}&	Android Malware Detection&	Input encoded to 32-dimension with positional encoding \& fed to transformer with 2 attention heads resulting \textbf{small model size} (1.4MB), suitable for mobile&Transformer-based dynamic feature aggregation provided a\textbf{ lightweight yet powerful method} for detecting malware in real-time \textbf{suitable for mobile devices}\\
 \hline
 
BERT [2021] ~\cite{rahali2021malbert}&	Android malware detection&	\textbf{Fine-tuned} BERT on malware source code with hyperparameter tuning&	Achieved 97.6\% accuracy in binary classification and 91.0\% accuracy in cross-category malware classification\\
\hline

\rowcolor{gray!10}
 Transformer in Multimodal Network [2021]~\cite{dehunting}&	Android malware detection&	Transformer Network (TN) encoder\textbf{ used for temporal feature extractor} from system call sequences in the dynamic analysis subnetwork (Chimera-D)&This adaptation \textbf{reduces noise and captures important temporal patterns }from system call data, leading to more accurate malware detection\\
 \hline

  Heterogeneous Temporal Graph Transformer (HTGT) [2021]~\cite{fan2021heterogeneous}&	Evolving Android malware detection&	HTGT integrates both spatial and temporal dependencies using \textbf{Heterogeneous Spatial Transformer(HST)} for spatial relations among different types of entities and \textbf{Temporal Transformer (TT)} to \textbf{joint model} the malware propagation and evolution through time&	The iterative interaction between HST and TT enhances the ability to \textbf{capture} both \textbf{malware propagation patterns and its evolutionary nature}\\
\hline

\rowcolor{gray!10}
BERT [2022]~\cite{ullah2022explainable}&	Android malware detection &	BERT was \textbf{fine-tuned} to extract text features from HTTP and TCP flows combined with image generated from network byte streams&\textbf{Combining textual and visual features} allows for a comprehensive analysis of both structural and behavioral malware indicators, \textbf{enhancing detection accuracy}\\
\hline

\textbf{SHERLOCK:} ViT with self-supervised learning [2022]~\cite{seneviratne2022self}	&Android malware detection &\textbf{Self-supervised} masked autoencoder for \textbf{pre-training}, where 75\% of the pixels in the image were masked and tasked to reconstruct from the unmasked 25\%, then \textbf{fine-tuned} for malware classification &Self-supervised learning approach was \textbf{effective in reducing overfitting}, especially with i\textbf{mbalanced classes} and pre-training enabled to learn robust representations\\
 \hline

 \rowcolor{gray!10}
MalBERTv2 (domain specific tokenization) [2023]~\cite{rahali2023malbertv2}&	Android Malware Identification&Introduces \textbf{pre-tokenization feature generator} which is trained from scratch on 85,000 Android APK samples and applies to \textbf{fine-tuned BERT}&Code-aware tokenization and fine-tuning proved \textbf{highly effective in multi-class classification}\\

\hline
BERT [2023]~\cite{saracino2023graph}	&Detection and categorization of Android malware&\textbf{Fine-tuned} BERT with data augmentation by manipulating existing sequences to generate new ones, helping to balance the dataset&Able to detect\textbf{ complex malware patterns}\\
\hline

\rowcolor{gray!10}
Vision Transformer (ViT) with attention map [2023]~\cite{jo2023malware}&	Malware detection and interperability&ViT adapted to take Android app DEX files, converting them into RGB images, generates \textbf{attention map} highlighting specific regions of contributions&Attention maps allows for \textbf{clear visual and textual explanations} of malware detection results\\
\hline

\end{tabular}
\end{adjustbox}
\end{table*}

\subsection{Performance Measures}
\label{performancemeasure}
Based on the observation of performance measures from the reviewed literature, we present our key insights on several aspects. 
\subsubsection{Accuracy Trends}
Aligned with the general trend of harnessing transformer models, we found that the performances are enhanced exceptionally well, achieving an accuracy upwards of 95\% in most cases. For example, the transformer with Multi-Head Attention and Bi-LSTM achieved a 99.63\% accuracy in Android malware detection tasks~\cite{chen2020android}. Similarly, the BERT with CNN-BiLSTM achieved a 99.39\% accuracy in cross-architecture IoT malware detection~\cite{hamad2021bertdeep}. These results highlight the strong performance of transformer models across malware detection and classification tasks, regardless of the specific system (like Android, IoT, Windows). 

\subsubsection{Accuracy Enhancement by Task-Specific Customization}
 Customized architectures that blend transformers with task-specific components (e.g., CNN, Bi-LSTM, graph-based layers) deliver significant improvements. The Directional Graph Transformer (DGT) for malware classification added graph embeddings and improved the accuracy by 37.5\% compared to standard baselines~\cite{moon2021directional}. Similarly, the Hierarchical Transformer for IoT malware detection leveraged internal structural information from network traffic to achieve 98\% accuracy ~\cite{hu2020exploit}. These adaptations demonstrate that task-specific customizations enhance both model precision and robustness in malware detection.

\subsubsection{Trade-Offs Between Accuracy and Computational Requirements}
While transformers deliver high accuracy, there is often a trade-off in terms of computational cost. Complex models such as transformer with Multi-Head Attention and Bi-LSTM demand more computational power to process intricate input data but yield superior accuracy~\cite{chen2020android}. Similarly, the Directional Graph Transformer (DGT), which uses control flow graphs (CFG), requires greater computational resources, but this added complexity results in enhanced detection capabilities~\cite{moon2021directional}. Thus, choosing a model depends on balancing accuracy with the available computational resources.

\subsubsection{Efficiency and Effectiveness of Transfer Learning}
Transfer learning is highly effective in malware tasks, significantly reducing the need for extensive training while maintaining high accuracy. Pre-trained models like BERT, GPT-2, XLNet, DistilBERT, ULMFiT, etc. excel when fine-tuned for malware tasks, with BERT achieving the accuracy of more than 90\% in malware detection and classification~\cite{bellante2021victory,ullah2022explainable, saracino2023graph, rahali2021malbert}. Such results show the strength of transfer learning in applying pre-trained language models to domain-specific learning tasks, yielding both time efficiency and performance improvements.

\subsection{Practical Perspectives}
To put some light upon the practical insights on the implementation of transformer models in real-world systems, we now discuss some of the examples from the literature.
\subsubsection{ViT4Mal for Resource-Constrained Environments}
Ravi et al.~\cite{ravi2023vit4mal} implemented a lightweight vision transformer (ViT4Mal) specifically optimized for resource-constrained environments like IoT devices. Their work addresses the challenges of computational complexity and memory usage while maintaining detection accuracy. This solution highlights the practical use of transformers in environments where computational power is limited -- a key real-world concern for IoT security. The study shows a 41x faster processing on FPGA boards compared to the original Vision Transformer (ViT), demonstrating how transformers can be tailored for low-resource environments.

\subsubsection{Heterogeneous Temporal Graph Transformer-HTGT}
Fan et al.~\cite{fan2021heterogeneous} developed the HTGT model, which integrates both spatial and temporal transformers to analyze Android malware evolution and propagation. This model was deployed in real-world mobile security solutions, protecting over 700 million mobile users. HTGT is designed to tackle the rapidly changing nature of Android malware.

\subsubsection{Galaxy Transformer for Complex Malware Analysis}
The Galaxy Transformer, proposed by Li et al.~\cite{li2021mad}, provides a unique architecture inspired by astronomical systems to model complex hierarchical structures. The transformer is arranged in star-plus and star-galaxy configurations to capture both basic and complex semantic structures in malware. This model effectively handles large datasets and complex sequences and was designed to address obfuscation techniques used by real-world malware, demonstrating how transformer models can enhance sophisticated malware detection.

\subsubsection{GPT-2 for Malware Evasion and Generation}
Another notable approach is based on the use of GPT-2 for generating adversarial malware examples. 
Hu et al.~\cite{hu2021single} fine-tuned GPT-2 to generate benign-looking perturbations that could bypass malware detection systems. This case highlights the use of transformer models not only for detection but also for adversarial testing, providing practical insights into how transformers can be used in a red-teaming context to assess and improve the robustness of security systems.

\subsubsection{Multi-Modal Transformers for Comprehension}
Multi-modal transformers, like those implemented by 
Ullah et al.~\cite{ullah2022explainable} for IoT malware detection, combine textual and visual representations to improve the accuracy of malware detection. Their system leverages BERT for textual feature extraction and converts malware binaries into images for CNN-based analysis, demonstrating the practical benefit of multi-modal data fusion for enhancing detection capabilities in real-world settings.

\end{document}